\documentclass[12pt]{article}
\usepackage[dvipsnames]{xcolor}
\usepackage{amsmath} 
\usepackage{amssymb}
\usepackage{mathtools}
\usepackage{physics}
\usepackage{bm}
\usepackage[margin =27truemm]{geometry}
 \usepackage{cite}
\pdfoutput=1
\usepackage[pdftex]{graphicx}

\newcommand{\ppp}{(+,+,+)}
\newcommand{\ppm}{(+,+,-)}
\newcommand{\pmm}{(+,-,-)}
\newcommand{\pmp}{(+,-,+)}
\newcommand{\mpp}{(-,+,+)}
\newcommand{\mpm}{(-,+,-)}
\newcommand{\mmm}{(-,-,-)}
\newcommand{\mmp}{(-,-,+)}
 
\begin{document}
\setlength{\baselineskip}{0.6cm}

\begin{titlepage}
\begin{flushright}
NITEP 240
\end{flushright}

\vspace*{10mm}%

\begin{center}{\Large\bf
SU(7) Grand Gauge-Higgs Unification 
}
\end{center}
\vspace*{10mm}
\begin{center}
{\large Yuzuho Komori}$^{a}$ and
{\large Nobuhito Maru}$^{a,b}$ 
\end{center}
\vspace*{0.2cm}
\begin{center}
${}^{a}${\it
Department of Physics, Osaka Metropolitan University, \\
Osaka 558-8585, Japan}
\\
${}^{b}${\it Nambu Yoichiro Institute of Theoretical and Experimental Physics (NITEP), \\
Osaka Metropolitan University,
Osaka 558-8585, Japan}
\end{center}

\vspace*{20mm}


\begin{abstract}
We propose a model of six dimensional $SU(7)$ grand gauge-Higgs unification compactified on $S^1/Z_2 \times S^1/Z_2$, 
which is a six dimensional extension of five dimensional $SU(4)$ gauge-Higgs unification 
predicting the weak mixing angle $\sin^2 \theta_W=1/4$ at the compactification scale. 
We investigate whether the correct pattern of electroweak symmetry breaking 
is realized in our model. 
We find some solutions providing the electroweak symmetry breaking and a realistic Higgs mass. 
Corresponding to each solution, the compactification scale of the fifth dimension is predicted. 
Furthermore, we evaluate the weak mixing angle at the weak scale by using one-loop renormalization group equation 
and find  it to be in good agreement with the experimental data.  
\end{abstract}

\end{titlepage}

\section{Introduction}
Gauge-Higgs Unification (GHU) is one of the fascinating scenarios beyond the Standard Model (SM) of particle physics, 
in which the Higgs field in the SM is unified as the extra spatial component of the higher dimensional gauge field 
\cite{Manton, Fairlie, Hosotani1, Hosotani2}.  
In this scenario, the Higgs mass is finite (at least one-loop level) by the gauge symmetry regardless of its nonrenormalizability 
\cite{HIL, SSS, ABQ, MY, HMTY, LMH, HLM} and it cannot be so much large comparing to the weak scale.   
Therefore, the gauge hierarchy problem is solved. 
Since the strong interaction is not included in GHU 
and the gauge hierarchy problem is originally discussed in the grand unified theory (GUT), 
it is very natural to consider the extension of GHU to the GUT, which we call the grand GHU (GGHU). 
GGHU has been much attracted and there has been many researches on GGHU in various context 
\cite{LM, KTY, HosotaniGGHU, MYT, ABBGW, MN}. 

In this paper, we construct a six dimensional (6D) $SU(7)$ GGHU model, 
which is an extension of five dimensional (5D) $SU(4)$ GHU model \cite{HLM}. 
The reason of considering 5D $SU(4)$ GHU model is as follows. 
In this model, the SM Higgs field is identified in the fifth component of the 5D gauge field 
and embedded into the two-rank anti-symmetric tensor (equivalently, anti-fundamental) representation of $SU(3)$, 
which is a subgroup of $SU(4)$. 
As a result, the weak mixing angle is predicted to be $\sin^2 \theta_W =1/4$ at the compactification scale. 
In GHU, the compactification scale is related to the weak scale and cannot be so large compared to the weak scale, 
typically ${\cal O}(1)$ TeV. 
Therefore, the prediction of the weak mixing angle $\sin^2 \theta_W =1/4$ at the compactification scale 
is in good agreement with the experimental data $\sin^2 \theta_W \sim 0.23$ at the weak scale as a leading order approximation. 
Thus, we consider to construct a GGHU model involving the 5D $SU(4)$ GHU 
while keeping the prediction of the weak mixing angle. 
In order to realize such a strategy, we propose a 6D $SU(7)$ GGHU model with the compactification on the orbifold 
$S^1/Z_2 \times S^1/Z_2$. 
After explaining the basic model properties, we calculate one-loop effective potential of SM Higgs field 
since the Higgs potential at tree level is forbidden by the gauge symmetry. 
Then, we investigate possibilities that the correct pattern of electroweak symmetry breaking 
and a realistic Higgs mass are realized in our model. 
Viable solutions will be found by introducing some 6D fermions additionally 
and the compactification scale of the fifth dimension is predicted. 
Furthermore, one-loop renormalization group equation (RGE) effects for the weak mixing angle 
between the compactification scale and the weak scale are calculated. 
The weak mixing angle at the weak scale is found to be in good agreement with the experimental data. 

This paper is organized as follows. 
In section 2, we introduce our model. 
In section 3, one-loop effective potential for the SM Higgs field are calculated 
to investigate the electroweak phase transition in the next section. 
The electroweak symmetry breaking analysis is performed in section 4. 
The weak mixing angle at the weak scale is evaluated by using one-loop RGE in section 5. 
Section 6 is a conclusion of our paper. 
In appendix A, $SU(7)$ generators are listed. 
In appendix B, the calculation of the Kaluza-Klein (KK) mass spectrum for 4D gauge boson and $A_5, A_6$ 
are described in some detail, which is needed to calculate one-loop effective potential.   
In appendix C, the calculation of one-loop effective potential is explained.   


\section{$SU(7)$ GGHU Model}
\subsection{Gauge sector}
Let us consider a model of six dimensional (6D) $SU(7)$ grand gauge-Higgs unification. 
First, the action of the gauge field is given. 
\begin{align}
  &S_{{\rm gauge}} = \int d^4 x \int dx_5 \int dx_6
  \left\{
  -\frac{1}{2}{\rm Tr}(F_{MN}F^{MN})\right\}, \\
  &F_{MN} = \partial_M A_N - \partial_N A_M -ig[A_M, A_N],
  \quad A_M = \sum_{a=1}^{48} A_M^a T^a,
\end{align}
where $T^a~(a=1,\cdots, 48)$ is $SU(7)$ generators. 
$A_M = A_\mu, A_5, A_6$ is a 6D gauge field, in which $M=0,1,2,3,5,6$ is a 6D index and $\mu=0,1,2,3$ is a 4D one. 
$g$ is a 6D gauge coupling constant. 
In order to obtain a 4D effective theory, we compactify each extra space $x_{5,6}$ on the orbifold $S^1/Z_2$. 
The sixth space is first compactified and then the fifth space is compactified, 
where we assume the radius $R_{5,6}$ to be $R_5 \gg R_6$ in this paper. 
The compactification scale of the sixth dimension $1/R_6$ is an arbitrary parameter, 
which is the Planck scale for instance. 

$Z_2$ parities for the gauge field around the fixed points $x_6=0, \pi R_6$ 
and $x_5 =0, \pi R_5$ are chosen as follows.\footnote{The boundary condition for $S^1$ is not independent 
and is shown to be $P_6^2$ and $P_5P'_5$.} 
\begin{align}
  A_{\mu,5} (x,x_5,-x_6) &= P_6 A_{\mu,5} (x,x_5,x_6) P_6, \\
  A_6 (x,x_5,-x_6) &= -P_6 A_6 (x,x_5,x_6) P_6, \\
  A_{\mu,5} (x,x_5,\pi R_6 -x_6) &= P_6 A_{\mu,5} (x,x_5,\pi R_6 +x_6) P_6, \\
  A_6 (x,x_5,\pi R_6 -x_6) &= -P_6 A_6 (x,x_5,\pi R_6 +x_6) P_6 
\end{align}
and 
\begin{align}
  A_{\mu,6} (x,-x_5,x_6) &= P_5 A_{\mu,6 }(x,x_5,x_6) P_5, \\
  A_5 (x,-x_5,x_6) &= -P_5 A_5 (x,x_5,x_6) P_5, \\
  A_{\mu,6} (x,\pi R_5 -x_5,x_6) &= P_5' A_{\mu,6} (x,\pi R_5 +x_5,x_6) P_5', \\
  A_5 (x,\pi R_5 -x_5,x_6) &= -P_5' A_5 (x,\pi R_5 +x_5,x_6) P_5', 
\end{align}
where the parity matrices $P_6, P_5, P_5'$ are explicitly given as
\begin{align}
\label{Z_2parity6}
  &P_6 = {\rm diag}(+, +, +, -, -, -, -), \\
  &P_5 = {\rm diag}(+, +, +, +, +, -, -), 
  \label {Z_2parity5}\\
  &P_5' = {\rm diag}(+, +, +, -, -, -, +).
\label{Z_2parity5'}
\end{align}
As will be seen below, these $Z_2$ parities are taken such that the SM gauge group will be unbroken 
and the SM Higgs field will remain to be massless. 
Note also that the $Z_2$ parities are commonly taken with respect to the fixed points in $x_6$ direction. 

Expressing the $Z_2$ parities of the components in the gauge field in a matrix form, 
we obtain
\begin{equation}
  A_\mu = 
  \left(
    \begin{array}{ccc|cc|c|c}
    \ppp & \ppp & \ppp & \mpm & \mpm & \mmm & \mmp \\
    \ppp & \ppp & \ppp & \mpm & \mpm & \mmm & \mmp \\
    \ppp & \ppp & \ppp & \mpm & \mpm & \mmm & \mmp \\
    \hline
    \mpm & \mpm & \mpm & \ppp & \ppp & \pmp & \pmm \\
    \mpm & \mpm & \mpm & \ppp & \ppp & \pmp & \pmm \\
    \hline
    \mmm & \mmm & \mmm & \pmp & \pmp & \ppp & \ppm \\
    \hline
    \mmp & \mmp & \mmp & \pmm & \pmm & \ppm & \ppp
    \end{array}
  \right), 
\end{equation}
\begin{equation}
  A_5 = 
  \left(
    \begin{array}{ccc|cc|c|c}
    \pmm & \pmm & \pmm & \mmp & \mmp & \mpp & \mpm \\
    \pmm & \pmm & \pmm & \mmp & \mmp & \mpp & \mpm \\
    \pmm & \pmm & \pmm & \mmp & \mmp & \mpp & \mpm \\
    \hline
    \mmp & \mmp & \mmp & \pmm & \pmm & \ppm & \ppp \\
    \mmp & \mmp & \mmp & \pmm & \pmm & \ppm & \ppp \\
    \hline
    \mpp & \mpp & \mpp & \ppm & \ppm & \pmm & \pmp \\
    \hline
    \mpm & \mpm & \mpm & \ppp & \ppp & \pmp & \pmm
    \end{array}
  \right),
\end{equation}
\begin{equation}
  A_6 = 
  \left(
    \begin{array}{ccc|cc|c|c}
    \mpp & \mpp & \mpp & \ppm & \ppm & \pmm & \pmp \\
    \mpp & \mpp & \mpp & \ppm & \ppm & \pmm & \pmp \\
    \mpp & \mpp & \mpp & \ppm & \ppm & \pmm & \pmp \\
    \hline
    \ppm & \ppm & \ppm & \mpp & \mpp & \mmp & \mmm \\
    \ppm & \ppm & \ppm & \mpp & \mpp & \mmp & \mmm \\
    \hline
    \pmm & \pmm & \pmm & \mmp & \mmp & \mpp & \mpm \\
    \hline
    \pmp & \pmp & \pmp & \mmm & \mmm & \mpm & \mpp
    \end{array}
  \right). 
\end{equation}
$Z_2$ parities for each component are denoted as $(P_6, P_5, P'_5)$. 
It is straightforward to expand the 6D gauge field consistently with the combinations of various $Z_2$ parities as follows. 
\begin{align}
  A^{a\ppp}_{M}(x,x_5,x_6)
  &=\frac{1}{2\pi\sqrt{R_5 R_6}}
  \left[
  A_{M}^{a(0,0)}(x)
  + \sqrt{2} \sum_{n=1}^\infty A_{M}^{a(n,0)}(x)\cos(\frac{n}{R_5}x_5) 
  \right. \nonumber \\
 &\left. \hspace*{-30mm}+ \sqrt{2} \sum_{m=1}^\infty A_{M}^{a(0,m)}(x)\cos(\frac{m}{R_6}x_6)
  + 2 \sum_{n=1}^\infty \sum_{m=1}^\infty
  A_{M}^{a(n,m)}(x)\cos(\frac{n}{R_5}x_5)\cos(\frac{m}{R_6}x_6)
  \right], \\
  A^{a\ppm}_{M}(x,x_5,x_6)
  &= \frac{1}{2\pi\sqrt{R_5 R_6}}
  \left[
  \sqrt{2} \sum_{n=0}^\infty
  A_{M}^{a(n,0)}(x)\cos(\frac{n+\frac{1}{2}}{R_5}x_5) 
  \right. \nonumber \\
  &\left. \hspace*{20mm} + 2 \sum_{n=0}^\infty \sum_{m=1}^\infty
  A_{M}^{a(n,m)}(x)\cos(\frac{n+\frac{1}{2}}{R_5}x_5)\cos(\frac{m}{R_6}x_6)
  \right], \\
  A^{a\pmp}_{M}(x,x_5,x_6)
  &= \frac{1}{2\pi\sqrt{R_5 R_6}}
  \left[
  \sqrt{2} \sum_{n=0}^\infty
  A_{M}^{a(n,0)}(x)\sin(\frac{n+\frac{1}{2}}{R_5}x_5) 
  \right. \nonumber \\
  &\left. \hspace*{20mm} + 2 \sum_{n=0}^\infty \sum_{m=1}^\infty
  A_{M}^{a(n,m)}(x)\sin(\frac{n+\frac{1}{2}}{R_5}x_5)\cos(\frac{m}{R_6}x_6)
  \right], \\
  A^{a\pmm}_{M}(x,x_5,x_6)
  &= \frac{1}{2\pi\sqrt{R_5 R_6}}
  \left[
  \sqrt{2} \sum_{n=1}^\infty
  A_{M}^{a(n,0)}(x)\sin(\frac{n}{R_5}x_5) 
  \right.  \nonumber \\
  &\left. \hspace*{20mm} + 2 \sum_{n=1}^\infty \sum_{m=1}^\infty
  A_{M}^{a(n,m)}(x)\sin(\frac{n}{R_5}x_5)\cos(\frac{m}{R_6}x_6)
  \right], \\
  A^{a\mpp}_{M}(x,x_5,x_6)
  &= \frac{1}{2\pi\sqrt{R_5 R_6}}
  \left[
  \sqrt{2} \sum_{m=1}^\infty
  A_{M}^{a(0,m)}(x)\sin(\frac{m}{R_6}x_6) 
  \right. \nonumber \\
  &\left. \hspace*{20mm} + 2 \sum_{n=1}^\infty \sum_{m=1}^\infty
  A_{M}^{a(n,m)}(x)\cos(\frac{n}{R_5}x_5)\sin(\frac{m}{R_6}x_6)
  \right], \\
  A^{a\mpm}_{M}(x,x_5,x_6)
  &= \frac{1}{\pi\sqrt{R_5 R_6}}\sum_{n=0}^\infty \sum_{m=1}^\infty 
  A_{M}^{a(n,m)}(x)\cos(\frac{n+\frac{1}{2}}{R_5}x_5)\sin(\frac{m}{R_6}x_6), \\
  A^{a\mmp}_{M}(x,x_5,x_6)
  &= \frac{1}{\pi\sqrt{R_5 R_6}}\sum_{n=0}^\infty \sum_{m=1}^\infty 
  A_{M}^{a(n,m)}(x)\sin(\frac{n+\frac{1}{2}}{R_5}x_5)\sin(\frac{m}{R_6}x_6), \\
  A^{a\mmm}_{M}(x,x_5,x_6)
  &= \frac{1}{\pi\sqrt{R_5 R_6}}\sum_{n=1}^\infty \sum_{m=1}^\infty 
  A_{M}^{a(n,m)}(x)\sin(\frac{n}{R_5}x_5)\sin(\frac{m}{R_6}x_6).
\end{align}
In these expressions, only the $\ppp$ components are special in that the fields with $Z_2$ parity $\ppp$ have 4D massless fields. 
We can see from $A_\mu$ that the gauge bosons of $SU(3)_C \times SU(2)_L \times U(1)_Y \times U(1)_X \times U(1)$ are massless, 
which means a gauge symmetry breaking by the compactification as 
\begin{align}
  SU(7)
  &\xrightarrow{S^1/Z_2~{\rm in}~x_6}
  SU(3)_C \times SU(4) \times U(1) \\
  &\xrightarrow{S^1/Z_2~{\rm in}~x_5}
  SU(3)_C \times SU(2)_L \times U(1)_Y \times U(1)_X \times U(1).
\end{align}
The diagonal generators of $SU(7)$ corresponding to 
the third component of the weak isospin $SU(2)_L$, $U(1)_Y$, $U(1)_X$ and $U(1)$ 
are given as follows. 
\begin{align}
&T^{15} = \frac{1}{2} {\rm diag}(0,0,0,1,-1,0,0), \\
&T^{24} = \frac{1}{2\sqrt{3}} {\rm diag}(0,0,0,1,1, -2,0), \\ 
&T^{35} = \frac{1}{2\sqrt{6}} {\rm diag}(0,0,0,1,1,1,-3), \\
&T^{48} = \frac{1}{2\sqrt{42}} {\rm diag}(4,4,4,-3,-3,-3,-3). 
\end{align}
The electromagnetic charge generator is given by the linear combination of $T^{15}$ and $T^{24}$, 
\begin{align}
Q_{em} = T^{15} + \sqrt{3} T^{24} ={\rm diag}(0,0,0,1,0,-1,0). 
\end{align}

In the first step of the compactification in the sixth space, 
we see that our model is indeed extended so as to have an $SU(4)$ as a subgroup. 
As for the extra $U(1)$ symmetries absent in the SM after the compactification in the fifth space, 
we can always give masses for these $U(1)$ gauge bosons by introducing the $U(1)$ charged 4D scalar fields 
and the potential similar to the Higgs potential in the SM by hand at the fixed points.  

Furthermore, we can see that the SM Higgs field $H$ is embedded in $A_5$ zero mode as
\begin{equation}
  A_5^{(0, 0)} =
  \left(
    \begin{array}{ccccccc}
      0 & 0 & 0 & 0 & 0 & 0 & 0 \\
      0 & 0 & 0 & 0 & 0 & 0 & 0 \\
      0 & 0 & 0 & 0 & 0 & 0 & 0 \\
      0 & 0 & 0 & 0 & 0 & 0 & \phi^+ \\
      0 & 0 & 0 & 0 & 0 & 0 & \phi^0 \\
      0 & 0 & 0 & 0 & 0 & 0 & 0 \\
      0 & 0 & 0 & \phi^- & \phi^{0*} & 0 & 0 \\
    \end{array}
  \right), \quad
  H =
  \left(
    \begin{array}{c}
      \phi^+ \\
      \phi^0
    \end{array}
  \right),
\end{equation}
where $\phi^\pm, \phi^0$ are obviously the charged and neutral scalar components of the Higgs field. 
Note that the neutral component of the Higgs field corresponds to the 44-th $SU(7)$ generator $T^{44}$ (see in Appendix A). 
As for $A_6$, we have no massless fields since there is no components with $Z_2$ parity $\ppp$. 
From these observations, the SM gauge bosons and the Higgs field are originated from the six dimensional gauge boson as massless fields. 

\subsection{Fermion sector}
Next, the action of the fermion in the 6D bulk is 
\begin{align}
S_{{\rm fermion}} = \int d^4 x \int dx_5 \int dx_6
\overline{\psi} i \Gamma^M D_M \psi, 
\end{align}
where the 6D gamma matrices $\Gamma^M$ are given by
\begin{equation}
  \Gamma^\mu
  =\left(
    \begin{array}{cc}
      \gamma^\mu & \\
      & \gamma^\mu
    \end{array}
  \right), \quad
  \Gamma^5
  =\left(
    \begin{array}{cc}
      & i\gamma^5\\
      i\gamma^5 &
    \end{array}
  \right), \quad
  \Gamma^6
  =\left(
    \begin{array}{cc}
      & \gamma^5 \\
      -\gamma^5 \\
    \end{array}
  \right) \quad
  (\mu=0,1,2,3). 
\end{equation}
$\gamma^\mu, \gamma^5$ are 4D gamma matrices. 
$\gamma_5$ is defined as $\gamma^5 = i \gamma^0 \gamma^1 \gamma^2 \gamma^3$, 
whose eigenvalues are $\pm 1$ specifying the 4D chirality. 

6D, 5D Dirac ($\psi, \psi_\pm$) and 6D, 4D Weyl fermions ($\psi_\pm, \psi_{L, R}$)  are defined as follows. 
\begin{align}
\psi_\pm = \frac{1 \pm \Gamma^7}{2} \psi, \qquad
\psi_{L} = \frac{1 - \gamma^5}{2} \psi_\pm,  \qquad 
\psi_{R} = \frac{1 + \gamma^5}{2} \psi_\pm, 
\end{align} 
where $\Gamma^7$ are 6D gamma matrix defined as 
\begin{align}
\Gamma^7= \Gamma^0 \Gamma^1 \Gamma^2 \Gamma^3 \Gamma^5 \Gamma^6, 
\end{align}
whose eigenvalues $\pm 1$ specifying 6D chirality. 
 
The $Z_2$ parity assignments for fermions are given as follows. 
\begin{align}
\psi_\pm(x_5, -x_6) &= \xi_R P_6(R) \Gamma^7 \psi_\pm(x_5, x_6), \\
\psi_\pm(x_5, \pi R_6 -x_6) &= \xi_R P_6(R) \Gamma^7 \psi_\pm(x_5, \pi R_6 + x_6), \\
\psi_{L, R}(-x_5, x_6) &= - \eta_R P_5(R) \gamma^5 \psi_{L, R}(x_5, x_6), \\
\psi_{L, R}(\pi R_5-x_5, x_6) &= - \eta'_R P'_5(R) \gamma^5 \psi_{L, R}(\pi R_5+x_5, x_6), 
 \end{align}
where $\xi_R, \eta_R, \eta_R = \pm1$ are undetermined parities depending on the representations $R$.
$P_6(R), P_5(R), P'_5(R)$ also depends on the representations and constructed by the tensor product of parity matrices 
$P_6, P_5, P'_5$ given by (\ref{Z_2parity6}), (\ref{Z_2parity5}) and(\ref{Z_2parity5'}).

Now, we analyze how the quarks and leptons in the SM are embedded in our model.
First, we consider the fundamental representation ${\bf 7}$ of $SU(7)$ and the decomposition into the irreducible ones 
under $SU(3)_C \times SU(2)_L \times U(1)_Y$.
\begin{align}
{\bf 7} &=  ({\bf 3}, {\bf 1})_{0}^{(\xi_7 \Gamma^7, - \eta_7 \gamma^5, - \eta'_7 \gamma^5)}
 \oplus ({\bf 1}, {\bf 2})_{1/2}^{(-\xi_7 \Gamma^7, - \eta_7 \gamma^5, \eta'_7 \gamma^5)}
 \nonumber \\
&\oplus ({\bf 1}, {\bf 1})_{0}^{(-\xi_7 \Gamma^7, \eta_7 \gamma^5, -\eta'_7 \gamma^5)}
\oplus ({\bf 1}, {\bf 1})_{-1}^{(-\xi_7 \Gamma^7, \eta_7 \gamma^5,  \eta'_7 \gamma^5)}
\label{7}
\end{align}
where the numbers in the parentheses are the representations of $SU(3)_C$ and $SU(2)_L$. 
The numbers in the subscript represent the hypercharges. 
The corresponding $Z_2$ parities of $(\xi_R P_6(R) \Gamma^7, -\eta_R P_5(R) \gamma^5, -\eta_R' P'_5(R) \gamma^5)$ 
are specified in the superscript. 
In these expressions, we have no SM left-handed lepton doublet although the SM right-handed electron is present.  

Next, let us consider a two rank anti-symmetric tensor representation ${\bf 21}$, 
which is decomposed into the representations 
under $SU(3)_C \times SU(2)_L \times U(1)_Y$, 
\begin{align}
{\bf 21} &=  
(\overline{{\bf 3}}, {\bf 1})_{0}^{(\xi_{21} \Gamma^7, -\eta_{21} \gamma^5, - \eta'_{21} \gamma^5)}
\oplus 
({\bf 1}, {\bf 2})_{1/2}^{(\xi_{21} \Gamma^7, \eta_{21}  \gamma^5, \eta'_{21} \gamma^5)}
\nonumber \\
&\oplus 
({\bf 1}, {\bf 2})_{-1/2}^{(\xi_{21} \Gamma^7, \eta_{21} \gamma^5, - \eta'_{21} \gamma^5)}
\oplus 
({\bf 1}, {\bf 1})_{1}^{(\xi_{21} \Gamma^7, -\eta_{21} \gamma^5, - \eta'_{21} \gamma^5)}
\nonumber \\
&\oplus 
({\bf 1}, {\bf 1})_{-1}^{(\xi_{21} \Gamma^7, -\eta_{21} \gamma^5, \eta'_{21} \gamma^5)}
\oplus 
({\bf 3}, {\bf 2})_{1/2}^{(-\xi_{21} \Gamma^7, -\eta_{21} \gamma^5, \eta'_{21} \gamma^5)}
\nonumber \\
&\oplus 
({\bf 3}, {\bf 1})_{0}^{(-\xi_{21} \Gamma^7, \eta_{21} \gamma^5, -\eta'_{21} \gamma^5)}
\oplus 
({\bf 3}, {\bf 1})_{-1}^{(-\xi_{21} \Gamma^7, \eta_{21} \gamma^5, \eta'_{21} \gamma^5)}. 
\end{align}
If we fix the parity of the left-handed lepton doublet $({\bf 1}, {\bf 2})_{L}~(\gamma^5=-1)$ as 
\begin{align}
(\xi_{21} \Gamma^7, -\eta_{21}, \eta'_{21})=(+,+,+)
 \end{align}
by hand, 
the right-handed electron $({\bf 1}, {\bf 1})_{-1}$ is realized as a massless zero mode. 
How the lepton doublet and singlet are embedded in ${\bf 21}$ representation can be shown in a matrix form as
\begin{align}
{\bf 21} \supset
\left(
\begin{array}{ccc|cc|c|c}
 & & & & & & \\
 & 3\times 3 & & & & & \\
 & & & & & & \\
 \hline
 & & & & & \nu_L & \\
 & & & & & e_L &  \\
 \hline
 & & & -\nu_L & -e_L & & e_R \\
 \hline
 & & & &  & -e_R & \\
 \end{array}
\right). 
\end{align}
%

The charged lepton Yukawa coupling can be obtained from the 6D gauge interaction of ${\bf 21}$ 
since the lepton doublet $({\bf 1}, {\bf 2})_L$ and the right-handed electron $({\bf 1}, {\bf 1})_R$ are included 
in the same $SU(7)$ multiplet. 
Noting the covariant derivative for ${\bf 21}$ are 
\begin{align}
D_M {\bf 21}^{ab} = \partial_M {\bf 21}^{ab} -ig A_M^C (T^C)^a_c {\bf 21}^{cb} -ig A_M^C (T^C)^b_c {\bf 21}^{ac},
\end{align}
where $T^C$ are the $SU(7)$ generators and $a,b,c = 1,\cdots 7, C=1,\cdots 48$.
In GHU, Yukawa coupling is generated from the gauge coupling as
\begin{align}
{\cal L}_{{\rm lepton~Yukawa}} 
&= \overline{{\bf 21}}^{ab} \left( 
g \langle A^{44}_5 \rangle \Gamma^5 (T^{44})_{bc} {\bf 21}^{ca} 
+ g \langle A^{44}_5 \rangle \Gamma^5 (T^{44})_{ac} {\bf 21}^{bc}
\right) \nonumber \\
&= 
2g \langle A^{44}_5  \rangle \left( 
\overline{{\bf 21}}^{65} \Gamma^5(T^{44})_{57} {\bf 21}^{76} 
+ \overline{{\bf 21}}^{67} \Gamma^5(T^{44})_{75} {\bf 21}^{56} 
\right) \nonumber \\
&= -g \langle A^{44}_5 \rangle \left( \overline{e}_R e_L + \bar{e}_L e_R \right), 
\end{align}
where 
$\Gamma^5 = i \gamma^5$. 
In the last equality, the chiral rotation ${\bf 21} \to e^{i \pi \gamma^5/4} {\bf 21}$ has been made 
to eliminate $\Gamma^5$. 

%

As for quarks, they cannot be embedded in the bulk field 
since the hypercharges in (\ref{7}) are integers  or half-integers, 
which implies that the fractional hypercharges of quarks cannot be obtained 
from the tensor product of the bulk fields in ${\bf 7}$ representation. 
Therefore, we introduce quarks $q_L, u_R, d_R$ as the 4D fermions on the brane at the origin in a compactified space $(0, 0)$, for instance. 
In this case, the quark Yukawa coupling can be introduced as follows. 
\begin{align}
{\cal L}_{\rm quark~Yukawa} = \delta(x_5) \delta(x_6) 
\left[
Y_u \overline{q}_L \tilde{A}_5 u_R + Y_d \overline{q}_L A_5 d_R
\right] + {\rm h.c.}
\end{align}
$\tilde{A}_5$ means an $SU(2)$ conjugated Higgs field. 
$Y_{u,d}$ are Yukawa coupling constants with mass dimension one. 
4D Yukawa coupling can be obtained by an integration over compactified space. 
\begin{align}
{\cal L}_{\rm quark~4D~Yukawa} &= 
\int dx_5 \int dx_6 \delta(x_5) \delta(x_6) 
\left[
Y_u \overline{q}_L \tilde{A}_5 u_R + Y_d \overline{q}_L A_5 d_R
\right] + {\rm h.c.} \nonumber \\
&= y_u \overline{q}_L \tilde{A}_5^{(0,0)} u_R + y_d \overline{q}_L A_5^{(0,0)} d_R
\end{align}
where $y_{u,d} = \frac{4\pi^2 R_5 R_6}{2\pi \sqrt{R_5R_6}}Y_{u,d}= 2\pi \sqrt{R_5 R_6}Y_{u,d}$ are 4D Yukawa couplings. 

\section{One-loop effective potential}
In this section, we study the electroweak symmetry breaking of $SU(7)$ GGHU model. 
In GHU or GGHU, the Higgs potential at tree level is forbidden by the gauge symmetry 
and we have to calculate the quantum corrections to Higgs potential. 
One-loop potential is calculated by Coleman-Weinberg potential, 
\begin{equation}
  V_{\rm eff}^{\rm 1-loop}
  = \frac{N}{(2 \pi)^2 R_5 R_6}
  \sum_{n,m} \int \frac{d^4 p_E}{(2 \pi)^4} \frac{(-1)^F}{2}
  \log(p_E^2 + m_{n,m}^2),
  \label{Veff}
\end{equation}
where $N$ is a physical degree of freedom for the fields running in the loop, 
$F$ is a fermion number, $p_E$ denotes a Euclidean momentum.  
$m_{n,m}$ is a KK mass of 6D bulk field, 
which depends on the vacuum expectation value (VEV) of the Higgs field in the background field method. 
In order to proceed the calculation, we have to obtain the KK mass spectrum of the fields. 

\subsection{One-loop potential from the gauge sector}

The KK mass terms come from the quadratic terms of the gauge kinetic terms. 
\begin{align}
  -\frac{1}{2}{\rm Tr}(F_{MN}F^{MN})
  & \supset
  -\frac{1}{2}{\rm Tr}
  (F_{5\mu}F^{5\mu}+F_{\mu 5}F^{\mu 5}
  +F_{6\mu}F^{6\mu}+F_{\mu 6}F^{\mu 6}
  +F_{56}F^{56}+F_{65}F^{65}) \notag \\
  &
  = \frac{1}{2}A_\mu^a(D_5 D^5 A^\mu)^a 
  + (\partial_\mu A_5)^a(D^5 A^\mu)^a - \frac{1}{2}(\partial_\mu A_5)^a(\partial^\mu A^5)^a \notag \\
  &\quad
  + \frac{1}{2}A_\mu^a(D_6 D^6 A^\mu)^a 
  + (\partial_\mu A_6)^a(D^6 A^\mu)^a - \frac{1}{2}(\partial_\mu A_6)^a(\partial^\mu A^6)^a \notag \\
  &\quad
  + \frac{1}{2}A_6^a(D_5 D^5 A^6)^a 
  + (\partial_6 A_5)^a(D^5 A^6)^a - \frac{1}{2}(\partial_6 A_5)^a(\partial^6 A^5)^a.
\end{align}
 Adding the following gauge-fixing term
\begin{equation}
  \mathcal{L}_{GF} 
  = \frac{1}{2\xi}
  \left\{
    \partial_\mu A^\mu
    -\xi (D^5 A_5 + D^6 A_6)
  \right\}^2,
\end{equation}
the mixing terms between $A_\mu$ and $A_{5,6}$ are cancelled and we find the relevant terms 
for the KK masses\footnote{It corresponds to $\xi=0$ case.}. 
\begin{equation}
  \frac{1}{2}A_\mu^a(D_5 D^5 A^\mu)^a
  + \frac{1}{2}A_\mu^a(D_6 D^6 A^\mu)^a
  + \frac{1}{2}A_6^a(D_5 D^5 A^6)^a 
  + (\partial_6 A_5)^a(D^5 A^6)^a - \frac{1}{2}(\partial_6 A_5)^a(\partial^6 A^5)^a. 
\end{equation}
Let us first focus on the KK mass terms for the  4D gauge fields. 
\begin{align}
  &\frac{1}{2} A_\mu^a(D_5 D^5 A^\mu)^a
  +\frac{1}{2} A_\mu^a(D_6 D^6 A^\mu)^a \notag \\
  &\qquad
  = - \frac{1}{2} A_\mu^a 
  \left\{
  (I^{ab}\partial_5 + gf^{a\,44\,b} \ev{A_5^{44}})
  (I^{bc}\partial_5 + gf^{b\,44\,c} \ev{A_5^{44}})
  + I^{ac}\partial_6 \partial_6
  \right\}
  A^{\mu \, c} + \cdots,
\end{align}
where $I^{ab}$ is a 48 $\times$ 48 unit matrix and $f^{abc}$ is a structure constant of $SU(7)$. 
$\cdots$ means terms irrelevant to the KK masses. 
Nonvanishing structure constants with an index ``44" indicating Higgs field are listed below.  
\begin{align}
  f^{15 \, 44 \, 45} &= -\frac{1}{2}, \quad 
  f^{16 \, 44 \, 37} = \frac{1}{2},\quad
  f^{17 \, 44 \, 36} = -\frac{1}{2},\quad
  f^{18 \, 44 \, 39} = \frac{1}{2}, \quad
  f^{19 \, 44 \, 38} = -\frac{1}{2}, \notag \\
  f^{20 \, 44 \, 41} &= \frac{1}{2}, \quad
  f^{21 \, 44 \, 40} = -\frac{1}{2}, \quad
  f^{22 \, 44 \, 43} = \frac{1}{2}, \quad
  f^{23 \, 44 \, 42} = -\frac{1}{2}, \quad
  f^{24 \, 44 \, 45} = \frac{1}{2\sqrt{3}}, \notag \\
  f^{33 \, 44 \, 47} &= \frac{1}{2}, \quad
  f^{34 \, 44 \, 46} = \frac{1}{2}, \quad
  f^{35 \, 44 \, 45} = \sqrt{\frac{2}{3}}. 
\end{align}
Diagonalizing the mass matrix of the 4D gauge field, we find the KK mass eigenvalues as
\begin{align*}
  m_{n,m}^2=&
  \frac{\alpha^2}{R_5^2}, \
  \frac{\alpha^2}{4R_5^2} \times 2, 
  \frac{(n\pm\alpha)^2}{R_5^2}, \
  \frac{(n\pm\frac{\alpha}{2})^2}{R_5^2} \times 2, \
  \frac{(\tilde{n}+\frac{1}{2}\pm\frac{\alpha}{2})^2}{R_5^2} \times 2, \\
  &
  \frac{\alpha^2}{R_5^2}+\frac{m^2}{R_6^2}, \
  \left(\frac{(\frac{\alpha}{2})^2}{R_5^2}+\frac{m^2}{R_6^2}\right) \times 2, 
  \frac{(n\pm\alpha)^2}{R_5^2}+\frac{m^2}{R_6^2}, \
  \left(\frac{(n\pm\frac{\alpha}{2})^2}{R_5^2}+\frac{m^2}{R_6^2}\right) \times 2, \\
  & \left(\frac{(\tilde{n}+\frac{1}{2}\pm\frac{\alpha}{2})^2}{R_5^2}+\frac{m^2}{R_6^2}\right) \times 8, 
  \stepcounter{equation}\tag{\theequation}
\end{align*}
where $n, m =1, \cdots, \infty, \quad \tilde{n}=0, \cdots, \infty$. 
The VEV of Higgs field is denoted as $\ev{A_5^{44}} = \frac{\alpha}{g R_5}$. 
The number multiplied by the KK masses means the number of degeneracy.  

Substituting these KK mass spectrum into (\ref{Veff}) 
and setting $N=4$ for the physical degree of freedom for $A_\mu$ and $F=0$, 
we obtain the one-loop effective potential from the contributions of $A_\mu$.
\begin{align}
  V_{\rm eff}^\mu &= -\frac{1}{32 \pi^3} \sum_{n=1}^\infty
  \left\{\sum_{m=1}^\infty 
  \frac{4}{\left((\pi R_5 n)^2 + (\pi R_6 m)^2\right)^3}
  + \frac{2}{(\pi R_5 n)^6}
  + \frac{3}{4(\pi R_5 n)^5 R_6}\right\} \notag \\
  & \quad \hspace{60mm} \times
  \left\{\cos(2 \pi n \alpha) + 2 \cos(\pi n \alpha)
  + 2(-1)^n \cos(\pi n \alpha)\right\} \notag \\
  & \quad 
  -\frac{1}{32 \pi^3} \sum_{n=1}^\infty 
  \left\{\sum_{m=1}^\infty 
  \frac{4}{\left((\pi R_5 n)^2 + (\pi R_6 m)^2\right)^3}
  + \frac{2}{(\pi R_5 n)^6}
  - \frac{3}{4(\pi R_5 n)^5 R_6}\right\}
  6(-1)^n \cos(\pi n \alpha).
  \label{eq:veffmu}
\end{align}

From the decomposition of the adjoint representation of $SU(7)$ into the irreducible representations 
under $SU(3)_C \times SU(2)_L$ and $Z_2$ parities $(P_6, P_5, P'_5)$, 
\begin{align}
  \bm{48} 
  &=
  (\bm{8},\bm{1})^{\ppp} \oplus (\bm{1},\bm{3})^{\ppp} \oplus
  (\bm{1},\bm{1})^{\ppp} \oplus (\bm{1},\bm{1})^{\ppm} 
  \oplus (\bm{1},\bm{1})^{\ppm} \nonumber \\
  & \oplus (\bm{1},\bm{2})^{\pmp} \oplus
  (\bm{1},\bar{\bm{2}})^{\pmp} \oplus (\bm{1},\bm{2})^{\pmm} 
  \oplus
  (\bm{1},\bar{\bm{2}})^{\pmm} \oplus (\bm{3},\bar{\bm{2}})^{\mpm} \nonumber \\
  & \oplus
  (\bm{3},\bm{1})^{\mmm} \oplus (\bm{3},\bm{1})^{\mmp} 
  \oplus
  (\bar{\bm{3}},\bm{2})^{\mpm} \oplus (\bar{\bm{3}},\bm{1})^{\mmm} \oplus
  (\bar{\bm{3}},\bm{1})^{\mmp} \oplus (\bm{1},\bm{1})^{\ppp} \nonumber \\
  &\supset
  (\bm{1},\bm{3})^{\ppp} \oplus (\bm{1},\bm{2})^{\pmp} \oplus
  (\bm{1},\bar{\bm{2}})^{\pmp} \oplus (\bm{1},\bm{2})^{\pmm} 
  \oplus
  (\bm{1},\bar{\bm{2}})^{\pmm} \nonumber\\
  &\oplus (\bm{3},\bar{\bm{2}})^{\mpm} \oplus (\bar{\bm{3}},\bm{2})^{\mpm}
  \label{48},
\end{align}
the one-loop effective potential from the the gauge boson 
with the $SU(2)_L$ representation $r$ can be expressed in a compact form 
\begin{multline}
  V_{\rm eff}^g(N, r, P_6, P_5, P'_5) =
  -\frac{N}{128 \pi^3}
  \sum_{n=1}^\infty
  \left\{
    \sum_{m=1}^\infty \frac{4}{((\pi R_5 n)^2 + (\pi R_6 m)^2)^3}
    + \frac{2}{(\pi R_5 n)^6}
    + \frac{3 P_6}{4(\pi R_5 n)^5 R_6}
  \right\} \\
  \times
  (P_5 P_5')^n
  \cos((r - 1)\pi n \alpha). 
  \label{Veffformula}
\end{multline}
%
$N$ denotes the physical degrees of freedom running in the loop. 
We note that only the multiplets belonging to nontrivial $SU(2)_L$ representations coupling to the Higgs field 
are extracted in the two lines from the bottom of (\ref{48}) 
since an $SU(2)_L$ singlet field have no contribution to the one-loop effective potential. 

Next, we calculate the KK mass spectrum of the $A_5$ and $A_6$. 
The quadratic terms of $A_{5,6}$ relevant to the KK masses are extracted from the gauge kinetic terms as follows. 
\begin{align*}
  &\frac{1}{2}A_6^a(D_5 D^5 A^6)^a 
  + (\partial_6 A_5)^a(D^5 A^6)^a - \frac{1}{2}(\partial_6 A_5)^a(\partial^6 A^5)^a \\
  &\qquad
  = - \frac{1}{2} A_6^a 
  (I^{ab}\partial_5 + gf^{a\,44\,b} \ev{A_5^{44}})
  (I^{bc}\partial_5 + gf^{b\,44\,c} \ev{A_5^{44}})
  A^{6 \, c} \\
  &\qquad \quad
  + \frac{1}{2} A_5^a 
  (I^{ac} \partial_5 \partial_6 + g f^{a\,44\,c}\ev{A_5^{44}} \partial_6) A^{6\,c}
  + \frac{1}{2} A_6^a 
  (I^{ac}\partial_5 \partial_6 + g f^{a\,44\,c}\ev{A_5^{44}}
  \partial_6) A^{5\,c} \\
  &\qquad \quad
  - \frac{1}{2}A_5^a(I^{ac} \partial_6 \partial_6) A^{5\,c} + \cdots
  \stepcounter{equation}\tag{\theequation}
\end{align*}
Similar to the 4D gauge field case, 
we find the KK masse spectrum of $A_{5,6}$ after diagonalizing the corresponding mass matrices, 
\begin{align}
  m_{n,m}^2=&
  \left(\frac{(\tilde{n}+\frac{1}{2}\pm\frac{\alpha}{2})^2}{R_5^2}\right) \times 6,\
  \frac{\alpha^2}{R_5^2}+\frac{m^2}{R_6^2}, \ 
  \left(\frac{\alpha^2}{4R_5^2}+\frac{m^2}{R_6^2}\right) \times 2,
  \frac{(n\pm\alpha)^2}{R_5^2}+\frac{m^2}{R_6^2}, \nonumber \\
   &\left(\frac{(n\pm\frac{\alpha}{2})^2}{R_5^2}+\frac{m^2}{R_6^2}\right) \times 2, \
  \left(\frac{(\tilde{n}+\frac{1}{2}\pm\frac{\alpha}{2})^2}{R_5^2}
  +\frac{m^2}{R_6^2}\right) \times 8,
  \stepcounter{equation}\tag{\theequation}
\end{align}
where $n, m=1, \cdots, \infty, \quad \tilde{n}=0, \cdots, \infty$. 

Substituting these KK mass spectrum into (\ref{Veff}) 
and setting $N=1$ for the physical degree of freedom for $A_{5,6}$\footnote{The reason why we take here 
the physical degrees of freedom to be $N=1$ not $N=2$ is the following. 
Some linear combination of $A_5$ and $A_6$ are absorbed as the longitudinal mode of the nonzero KK gauge fields 
similar to the Higgs mechanism and they are unphysical, but the other orthogonal linear combinations are remained to be as physical modes} 
and $F=0$, 
we obtain the one-loop effective potential from the contributions of $A_{5,6}$.
\begin{align}
  V_{\rm eff}^{56}
  &= -\frac{1}{128 \pi^3} \sum_{n=1}^\infty
  \left\{\sum_{m=1}^\infty 
  \frac{4}{\left((\pi R_5 n)^2 + (\pi R_6 m)^2\right)^3}
  + \frac{2}{(\pi R_5 n)^6}
  - \frac{3}{4(\pi R_5 n)^5 R_6}\right\} \notag \\
  & \quad \hspace{60mm} \times
  \left\{\cos(2 \pi n \alpha) + 2 \cos(\pi n \alpha)
  + 2(-1)^n \cos(\pi n \alpha)\right\} \notag \\
  & \quad
  -\frac{1}{128 \pi^3} \sum_{n=1}^\infty 
  \left\{\sum_{m=1}^\infty 
  \frac{4}{\left((\pi R_5 n)^2 + (\pi R_6 m)^2\right)^3}
  + \frac{2}{(\pi R_5 n)^6}
  + \frac{3}{4(\pi R_5 n)^5 R_6}\right\}
  6(-1)^n \cos(\pi n \alpha).
  \label{eq:veff56}
\end{align}
Combining the potential from $A_\mu$ \eqref{eq:veffmu} and that from $A_5,A_6$ \eqref{eq:veff56}, 
the one-loop effective potential from the 6D gauge boson is obtained. 
\begin{align}
  V_{\rm eff}^g
  &= V_{\rm eff}^\mu + V_{\rm eff}^{56} 
  \nonumber \\
  &= 
 \left[ V_{\rm eff}^g(4, 3, +,  +, +) + 2 V_{\rm eff}^g(4, 2, +, -, -) + 2 V_{\rm eff}^g(4, 2, +, -, +) + 6 V_{\rm eff}^g(4, 2, -, +, -) \right] \notag \\
 & \hspace*{10mm} 
 + \left[ V_{\rm eff}^g(1, 3, -, -, -) + 2 V_{\rm eff}^g(1, 2, -, +, +) + 2 V_{\rm eff}^g(1, 2, -, +, -) + 6 V_{\rm eff}^g(1, 2, +, -, +) \right] 
       \notag \\
  &=
  -\frac{1}{32 \pi^3} \sum_{n=1}^\infty
  \left\{\sum_{m=1}^\infty 
  \frac{4}{\left((\pi R_5 n)^2 + (\pi R_6 m)^2\right)^3}
  + \frac{2}{(\pi R_5 n)^6}
  + \frac{3}{4(\pi R_5 n)^5 R_6}\right\} 
  \nonumber \\
  & \quad \hspace{60mm} \times
  \left\{\cos(2 \pi n \alpha) + 2 \cos(\pi n \alpha)
  + 2(-1)^n \cos(\pi n \alpha)\right\} 
  \nonumber \\
  & \quad 
  -\frac{1}{32 \pi^3} \sum_{n=1}^\infty 
  \left\{\sum_{m=1}^\infty 
  \frac{4}{\left((\pi R_5 n)^2 + (\pi R_6 m)^2\right)^3}
  + \frac{2}{(\pi R_5 n)^6}
  - \frac{3}{4(\pi R_5 n)^5 R_6}\right\}
  6(-1)^n \cos(\pi n \alpha) 
  \nonumber \\
  & \quad 
  -\frac{1}{128 \pi^3} \sum_{n=1}^\infty
  \left\{\sum_{m=1}^\infty 
  \frac{4}{\left((\pi R_5 n)^2 + (\pi R_6 m)^2\right)^3}
  + \frac{2}{(\pi R_5 n)^6}
  - \frac{3}{4(\pi R_5 n)^5 R_6}\right\} 
  \nonumber \\
  & \quad \hspace{60mm} \times
  \left\{\cos(2 \pi n \alpha) + 2 \cos(\pi n \alpha)
  + 2(-1)^n \cos(\pi n \alpha)\right\} 
  \nonumber \\
  & \quad
  -\frac{1}{128 \pi^3} \sum_{n=1}^\infty 
  \left\{\sum_{m=1}^\infty 
  \frac{4}{\left((\pi R_5 n)^2 + (\pi R_6 m)^2\right)^3}
  + \frac{2}{(\pi R_5 n)^6}
  + \frac{3}{4(\pi R_5 n)^5 R_6}\right\}
  6(-1)^n \cos(\pi n \alpha).
\end{align}
Here the following part of the potential
\begin{equation}
  \sum_{m=1}^\infty 
  \frac{4}{\left((\pi n)^2 + (\frac{\pi m}{k})^2\right)^3}
  + \frac{2}{(\pi n)^6}
  +\frac{3 k P_6}{4(\pi n)^5},
\end{equation}
where 
\begin{equation}
  \frac{1/R_6}{1/R_5} = k
\end{equation}
is defined as the ratio between the two compactification scales, 
can be simplified as follows.  
Carrying out the summation with respect to $m$ explicitly, we find 
\begin{align}
  &\sum_{m=1}^\infty
  \frac{1}{((\pi n)^2 + (\frac{\pi m}{k})^2)^3} \\
  & \qquad 
  =\frac{-8 +3 \pi  k n \coth (\pi  k n)
  +3 \pi ^2 k^2 n^2 \text{csch}^2(\pi  k n)
  + 2 \pi ^3 k^3 n^3 \coth (\pi  k n) \text{csch}^2(\pi  k n)}{16 (\pi n)^6} \notag \\
  & \qquad  \sim
  \frac{-8 + 3 \pi  k n}{16 (\pi n)^6}, 
\end{align}
where $k \gg 1$ is assumed in our model 
and $\coth (x) \sim 1, \text{csch} (x) \sim 0$ are taken into account in the last line.  
Thus, we obtain
\begin{align*}
  \sum_{m=1}^\infty 
  \frac{4}{\left((\pi n)^2 + (\frac{\pi m}{k})^2\right)^3}
  + \frac{2}{(\pi n)^6}
  + \frac{3 k P_6}{4(\pi n)^5} 
  \sim 
  \frac{3(1 + P_6)k}{4 (\pi n)^5}. 
  \label{approx}
  \stepcounter{equation}\tag{\theequation}
\end{align*}
Using the approximation, the one-loop potential from the 6D gauge field can written as
\begin{align*}
  V_{\rm eff}^g
  & \sim
  -\frac{1}{32 \pi^3 R_5^6} \sum_{n=1}^\infty
  \frac{6k}{4 (\pi n)^5}
  \left\{\cos(2 \pi n \alpha) + 2 \cos(\pi n \alpha)
  + 2(-1)^n \cos(\pi n \alpha)\right\} \\
  &\quad
  -\frac{1}{128 \pi^3 R_5^6} \sum_{n=1}^\infty 
  \frac{6k}{4 (\pi n)^5}
  6(-1)^n \cos(\pi n \alpha)\\
  &= -\frac{3 k}{128 \pi^8 R_5^6} \sum_{n=1}^\infty
  \frac{1}{n^5}
  \left\{2\cos(2 \pi n \alpha) + 4 \cos(\pi n \alpha)
  + 7(-1)^n \cos(\pi n \alpha)\right\}. 
  \stepcounter{equation}\tag{\theequation}
  \label{eq:veffgauge}
\end{align*}

\subsection{One-loop potential from fermion sector}
In this section, we calculate the KK mass spectrum of massless fermions in the 6D bulk. 
As will be discussed later, we need to introduce the contributions of fermions to the effective potential 
since the correct pattern of electroweak symmetry breaking is not realized by the contribution from the gauge field only.  
In this paper, we consider fermions belonging to the fundamental, adjoint, two-rank and three-rank totally symmetric tensors of $SU(7)$. 
We note that these 6D fermions have nothing to do with the SM quarks and leptons, 
namely the SM fermions are not embedded in these $SU(7)$ multiplets.   

As discussed in the previous subsection, 
it is enough to find the decomposition pattern of $SU(7)$ representations into the representation 
under $(SU(3)_C, SU(2)_L)$ and $Z_2$ parities $(P_6, P_5, P'_5)$ 
to obtain the one-loop effective potential instead of diagonalizing the KK mass matrices. 
Since the fundamental and the adjoint representations have already been discussed in (\ref{7}) and (\ref{48}), 
the decomposition of the two-, and the three-rank totally symmetric tensors are listed below, 
where only the fields with the nontrivial representations of $SU(2)_L$ relevant to the one-loop potential are extracted. 
\begin{align}
{\bf 28}&\supset 
({\bf 3}, {\bf 2})_{-1/2}^{(-\xi_{28} \Gamma^7, -\eta_{28} \gamma^5,  \eta'_{28} \gamma^5)}
\oplus 
({\bf 1}, {\bf 3})_{-1}^{(\xi_{28} \Gamma^7, -\eta_{28} \gamma^5, -\eta'_{28} \gamma^5)}
\notag \\
&
\oplus 
({\bf 1}, {\bf 2})_{1/2}^{(\xi_{28} \Gamma^7, \eta_{28} \gamma^5, -\eta'_{28} \gamma^5)}
\oplus 
({\bf 1}, {\bf 2})_{-1/2}^{(\xi_{28} \Gamma^7, \eta_{28} \gamma^5, \eta'_{28} \gamma^5)}
\label{28}
\end{align}
\begin{align}
{\bf 84} &\supset
({\bf 6}, {\bf 2})_{1/2}^{(-\xi_{84} \Gamma^7, -\eta_{84} \gamma^5, \eta'_{84} \gamma^5)}
\oplus 
({\bf 3}, {\bf 3})_{1}^{(\xi_{84} \Gamma^7, -\eta_{84} \gamma^5, -\eta'_{84} \gamma^5)}
\notag \\
&\oplus 
({\bf 3}, {\bf 2})_{1/2}^{(\xi_{84} \Gamma^7, \eta_{84} \gamma^5, \eta'_{84} \gamma^5)}
\oplus
({\bf 3}, {\bf 2})_{-1/2}^{(\xi_{84} \Gamma^7, \eta_{84} \gamma^5, -\eta'_{84} \gamma^5)}
\notag \\
&\oplus 
({\bf 1}, {\bf 4})_{3/2}^{(-\xi_{84} \Gamma^7, -\eta_{84} \gamma^5, \eta'_{84} \gamma^5)}
\oplus 
({\bf 1}, {\bf 3})_{1}^{(-\xi_{84} \Gamma^7, \eta_{84} \gamma^5, -\eta'_{84} \gamma^5)}
\nonumber \\
&\oplus 
({\bf 1}, {\bf 3})_{0}^{(-\xi_{84} \Gamma^7, \eta_{84} \gamma^5, \eta'_{84} \gamma^5)}
\oplus 
({\bf 1}, {\bf 2})_{1/2}^{(-\xi_{84} \Gamma^7, -\eta_{84} \gamma^5, \eta'_{84} \gamma^5)}
\nonumber \\
&\oplus 
({\bf 1}, {\bf 2})_{-3/2}^{(-\xi_{84} \Gamma^7, -\eta_{84} \gamma^5, \eta'_{84} \gamma^5)}
\oplus 
({\bf 1}, {\bf 2})_{-1/2}^{(-\xi_{84} \Gamma^7, -\eta_{84} \gamma^5, -\eta'_{84} \gamma^5)}
\label{84}
\end{align}
In the case of fermion in the $SU(2)_L$ representation $r$ and $Z_2$ parities 
$(\xi_R P_6(R) \Gamma^7, -\eta_R P_5(R) \gamma^5, -\eta_R' P'_5(R) \gamma^5)$, 
one-loop effective potential is generalized in the following  
by taking into account the sign ambiguities $\xi_R, \eta_R, \eta'_R =\pm 1$ 
depending on the representation $R$ of $SU(7)$. 
\begin{align}
  &V_{\rm eff}^f(N_f, r, \xi_R P_6 \Gamma^7, \eta_R P_5 \gamma^5, \eta'_R P'_5 \gamma^5) \notag \\
  &\hspace*{10mm} =
  \frac{N_f}{128 \pi^3}
  \sum_{n=1}^\infty
  \left\{
    \sum_{m=1}^\infty \frac{4}{((\pi R_5 n)^2 + (\pi R_6 m)^2)^3}
    + \frac{2}{(\pi R_5 n)^6}
    + \frac{3 \xi_R P_6\Gamma^7}{4(\pi R_5 n)^5 R_6}
  \right\} \notag \\
  &\hspace*{20mm}\times
  (\eta_R P_5 \gamma^5 \eta'_R P_5' \gamma^5)^n
  \cos((r - 1)\pi n \alpha) \notag \\ 
 & \hspace*{10mm}\sim \frac{N_f k}{128 \pi^8 R_5^6}
  \sum_{n=1}^\infty
  \frac{3(1 + \xi_R P_6 \Gamma^7)}{4 n^5}
  (\eta_R P_5 \gamma^5 \eta'_R P_5' \gamma^5)^n
  \cos((r - 1)\pi n \alpha).
  \label{Vefffermion}
\end{align}
$N_f$ means physical degrees of freedom of fermions. 
Note that this potential represents that from the largest eigenvalues proportional to the Higgs VEV. 
For more than 3-rank symmetric tensor representation, we have to add the potential from corresponding smaller eigenvalues. 
For instance, we have to sum the potential from ${\bf 4}$ and ${\bf 2}$ for ${\bf 4}$.  

From the above expression, the one-loop effective potential for the 6D Dirac fermion can be found as follows. 
We note that a 6D Dirac fermion is decomposed into four 4D Weyl fermions $\psi_{+L}, \psi_{+R}, \psi_{-L}, \psi_{-R}$ 
depending on the eigenvalues of $(\Gamma^7, \gamma^5) = (\pm, \pm)$. 
Noticing also that the physical degrees of freedom of 4D Weyl fermion is two, 
we obtain the one-loop effective potential for the 6D Dirac fermion
\begin{align*}
  V_{\rm eff}^f (r,\xi_R P_6, \eta_R P_5, \eta_R' P_5')
  =& \frac{2k}{128 \pi^8 R_5^6}
  \sum_{n=1}^\infty
  \frac{3(1 + \xi_R P_6)}{4 n^5}
  (\eta_R P_5 \eta'_R P_5')^n
  \cos((r - 1)\pi n \alpha)\\
  &+
  \frac{2k}{128 \pi^8 R_5^6}
  \sum_{n=1}^\infty
  \frac{3(1 + \xi_R P_6)}{4 n^5}
  ((-1)^2 \eta_R P_5 \eta'_R P_5')^n
  \cos((r - 1)\pi n \alpha)\\
  &+
  \frac{2k}{128 \pi^8 R_5^6}
  \sum_{n=1}^\infty
  \frac{3(1 - \xi_R P_6)}{4 n^5}
  (\eta_R P_5 \eta'_R P_5')^n
  \cos((r - 1)\pi n \alpha)\\
  &+
  \frac{2k}{128 \pi^8 R_5^6}
  \sum_{n=1}^\infty
  \frac{3(1 - \xi_R P_6)}{4 n^5}
  ((-1)^2 \eta_R P_5 \eta'_R P_5')^n
  \cos((r - 1)\pi n \alpha)\\
  =&
  \frac{3 k}{64 \pi^8 R_5^6}
  \sum_{n=1}^\infty
  \frac{1}{n^5}
  (\eta_R P_5 \eta'_R P_5')^n
  \cos((r - 1)\pi n \alpha). 
  \stepcounter{equation}\tag{\theequation}
  \label{Veff6Dfermion}
\end{align*}
By applying the decompositions (\ref{7}), (\ref{48}), (\ref{28}) and (\ref{84}) to the formula (\ref{Veff6Dfermion}), 
it is straightforward to obtain one-loop effective potential from the fields 
in ${\bf 7}$, ${\bf 28}$, ${\bf 48}$ and ${\bf 84}$ representations with $\xi_R, \eta_R, \eta'_R$. 
\begin{align}
  V_{\rm eff}^f({\bf 7}) &= 
  V_{\rm eff}^f(2, -\xi_7, -\eta_7, \eta'_7), \\   
 V_{\rm eff}^f({\bf 28}) & =  
  V_{\rm eff}^f(3, \xi_{28}, -\eta_{28}, -\eta'_{28}) 
 + 3V_{\rm eff}^f(2, -\xi_{28}, -\eta_{28}, \eta'_{28}) \notag \\
 & + V_{\rm eff}^f(2, \xi_{28}, \eta_{28}, -\eta'_{28}) 
  + V_{\rm eff}^f(2, \xi_{28}, \eta_{28}, \eta'_{28}), \\
 V_{\rm eff}^f({\bf 48}) & =
V_{\rm eff}^f(3, \xi_{48}, -\eta_{48}, -\eta'_{48}) 
 + 6 V_{\rm eff}^f(2, -\xi_{48}, -\eta_{48}, \eta'_{48}) \notag \\
&  + 2V_{\rm eff}^f(2, \xi_{48}, \eta_{48}, -\eta'_{48}) 
  + 2V_{\rm eff}^f(2, \xi_{48}, \eta_{48}, \eta'_{48}), \\
 V_{\rm eff}^f({\bf 84}) & =
 V_{\rm eff}^f(4, -\xi_{84}, -\eta_{84}, \eta'_{84})  
  + V_{\rm eff}^f(3, -\xi_{84}, \eta_{84}, -\eta'_{84}) 
   + V_{\rm eff}^f(3, -\xi_{84}, \eta_{84}, \eta'_{84}) \notag \\
 &+ 3V_{\rm eff}^f(3, \xi_{84}, -\eta_{84}, -\eta'_{84}) 
 + 6V_{\rm eff}^f(2, -\xi_{84}, -\eta_{84}, \eta'_{84}) 
   + 3V_{\rm eff}^f(2, \xi_{84}, \eta_{84}, \eta'_{84}) \notag \\
  &+ 3V_{\rm eff}^f(2, \xi_{84}, \eta_{84}, -\eta'_{84})
  + 2 V_{\rm eff}^f(2, -\xi_{48}, -\eta_{48}, \eta'_{48}) 
  + V_{\rm eff}^f(2, -\xi_{48}, -\eta_{48}, -\eta'_{48}). 
\end{align}

Introducing the fermions like above, many unwanted massless fermions are remained as zero modes. 
In order to make these massless fermions heavy, we have to introduce the 4D fermion 
conjugate to the representations and charges of the massless fermions and their Dirac mass terms on the fixed points.  

\section{Electroweak symmetry breaking}
Now, it is ready to study the possibility of the electroweak symmetry breaking in our model. 
In GHU, we note that the order parameter of the electroweak symmetry is a VEV of Wilson line, 
which is nonlocal and a gauge invariant quantity,   
\begin{align}
  \ev{W}
  &=
  P \exp[ig \oint_{S^1}dx^5 \ev{A_5^{44}}] \notag \\
  &=
  \left(
    \begin{array}{ccccccc}
      1 & 0 & 0 & 0 & 0 & 0 & 0 \\
      0 & 1 & 0 & 0 & 0 & 0 & 0 \\
      0 & 0 & 1 & 0 & 0 & 0 & 0 \\
      0 & 0 & 0 & 1 & 0 & 0 & 0 \\
      0 & 0 & 0 & 0 & \cos(\pi \alpha) & 0 & i \sin(\pi \alpha) \\
      0 & 0 & 0 & 0 & 0 & 1 & 0 \\
      0 & 0 & 0 & 0 & i \sin(\pi \alpha) & 0 & \cos(\pi \alpha) \\
    \end{array}
  \right), 
\end{align}
and in order to obtain the information of the symmetry breaking pattern, 
we have to check that which of the generators commute with the VEV of Wilson line. 
Obviously, $SU(3)_C \times U(1)$ is unbroken 
since the $SU(3)_C$ generators $T^a(a=1, \cdots, 8)$ and the $U(1)$ generator $T^{48}$ 
are commutable with $\ev{W}$ irrespective of the VEV $\alpha_{\rm min}$ at the minimum of the potential. 
As for $SU(2)_L \times U(1)_Y \times U(1)_X$, 
there two the linear combination of $T^{15}, T^{24}$ and $T^{35}$
\begin{align}
  &U(1)_{em}: T^{15}+\sqrt{3}T^{24}={\rm diag}(0,0,0,1,0,-1,0), \\
  &U(1)': T^{15} - \frac{\sqrt{3}}{3}T^{24} + \frac{\sqrt{6}}{3} T^{35} = \frac{1}{2} {\rm diag}(0, 0, 0, 1, -1, 1, -1)
\end{align}
commute with $\ev{W}$ for $0<\alpha_{\rm min}<1$\footnote{In our model, 
we have only to consider $0<\alpha_{\rm min} \le 1$ 
since the potential is periodic between $\alpha=-1$ and $\alpha=1$ and symmetric with respect to $\alpha=0$. 
In the case of $\alpha=1$, extra $U(1)$ symmetries remains to be unbroken.}. 
In order to obtain the correct pattern of electroweak symmetry breaking 
$SU(2)_L \times U(1)_Y \to U(1)_{em}$, $U(1)'$ has to be spontaneously broken 
by introducing $U(1)'$ charged 4D scalar field on the fixed point and a symmetry breaking potential. 

We note that the potentials from the SM fermion contributions can be safely neglected. 
For the leptons to be embedded into the bulk fermion, the bulk fermion must be massive to generate lepton masses \cite{AHS}. 
The potential from such a massive bulk fermion has an extra factor $e^{-M\pi R} \sim e^{-{\cal O}(10)}$ ($M$: mass of the bulk fermion)
and is very suppressed comparing to that from the massless bulk fermion proportional to an enhancement factor $k~(\gg 1)$ 
as can be seen in (\ref{eq:veffgauge}) and (\ref{Veff6Dfermion}). 
For the quarks localized at the fixed point, 
their potential takes the form like 4D one-loop effective potential $m_q(\alpha)^4 \ln m_q(\alpha)^2$ 
which is proportional to $\alpha^4 \ll 1$ and has no enhancement factor $k$.  
From this observation, we do not consider the potential from the SM fermion contributions in this paper. 

\begin{figure}[h]
  \centering
  \includegraphics[width=60mm]{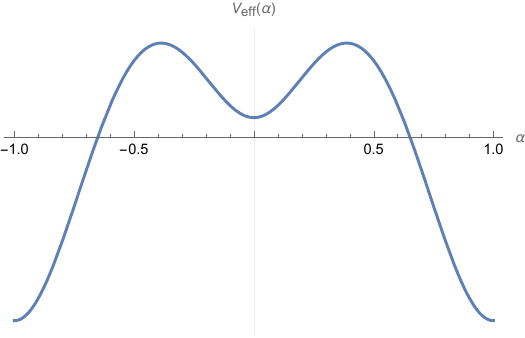}
  \caption{Higgs potential of the gauge field only ($\alpha_{\rm min}=1$).}
  \label{fig:g}
\end{figure}

We study whether the correct pattern of the electroweak symmetry breaking 
$SU(2)_L \times U(1)_Y \to U(1)_{em}$ is realized in our model by considering the various combinations of fermions. 
If we find such a symmetry breaking, we calculate Higgs mass and the compactification scales $1/R_{5, 6} $ as follows. 
The SM Higgs mass can be calculated from the second derivative of the one-loop effective potential, 
\begin{align}
  m_h^2 &= g^2 R_5^2
  \left.\pdv[2]{V_{\rm eff}}{\alpha}\right|_{\alpha=\alpha_{\rm min}} 
  = 4 \pi^2 g_4^2 R_5^3 R_6
  \left.\pdv[2]{V_{\rm eff}}{\alpha}\right|_{\alpha=\alpha_{\rm min}},
  \label{eq:mh}
\end{align}
where $g, g_4$ are the 6D, 4D gauge couplings, respectively and related by $g_4 = \frac{g}{2\pi\sqrt{R_5 R_6}}$. 
Noting that the W boson mass in 5D $SU(4)$ GHU is given by
\begin{align}
  m_W^2 = \frac{\alpha^2}{4R_5^2}, 
  \label{eq:wmass}
\end{align}
the compactificaiton scale of the fifth dimension $1/R_5$ can be evaluated at the potential minimum as
\begin{align}
  &\frac{1}{R_5} 
  = \frac{2 \times 80.4 \, {\rm [GeV]}}{\alpha_{\rm  min}}. 
  \label{eq:R5}
\end{align}
First, we show that the correct pattern of the electroweak symmetry breaking cannot be realized 
in no fermion case, that is, only the gauge field is present.  
The plot of the corresponding Higgs potential is shown in Fig.~\ref{fig:g}. 
As discussed above, the minimum $\alpha_{\rm min}=1$ is not our desired result. 

\begin{figure}[h]
  \begin{tabular}{cc}
  \begin{minipage}{.48\textwidth}
    \centering
    \includegraphics[width=65mm]{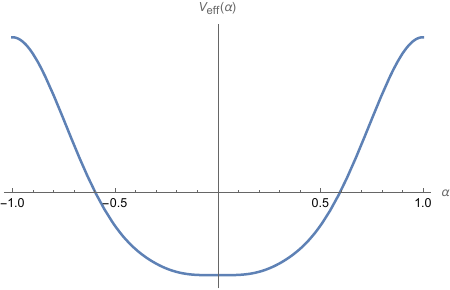}
    \label{fig:125}
  \end{minipage}
  \begin{minipage}{.48\textwidth}
    \centering
    \includegraphics[width=65mm]{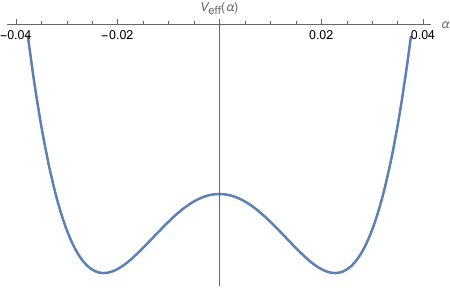}
    \label{fig:125_min}
  \end{minipage}
  \end{tabular}
  \caption{Plots of the Higgs potential in case (3) (left) and the same potential around the minimum (right).}
\label{case3}
\end{figure}

Next, we study the potential analysis by introducing additionally up to six fermions 
in ${\bf 7}, {\bf 28}, {\bf 48}$ and ${\bf 84}$ representations to improve the previous result. 
We investigate all possible combinations of various representations and 
the results satisfying the Higgs mass in a range 125-127 GeV 
and $\alpha_{\rm min} \sim{\cal O}(0.01)$\footnote{This condition is required through the relation $m_W =\alpha/(2R_5)$ 
where $1/R_5$ must be larger than ${\cal O}(1)$ TeV.} are summarized in Table \ref{result}. 
\begin{table}[h]
\begin{center}
\begin{tabular}{c|l|c|c|c}
& Fermion content & $\alpha_{\rm min}$ & $m_h$ (GeV) & $1/R_5$ (TeV) \\
\hline
(1) &${\bf 28}^{(+,-)}+4\times {\bf 84}^{(+,+)}$ & 0.043 & 126.8 & 3.8 \\
(2) & ${\bf 28}^{(+,+)} + 4\times {\bf 84}^{(+,+)}$ & 0.081 & 125.5 & 2.0 \\
(3) &${\bf 28}^{(+,+)}+ 3 \times {\bf 48}^{(+,+)} +2 \times {\bf 84}^{(+,+)}$ & 0.021 & 125.1 & 7.5 \\
(4) &${\bf 7}^{(+,-)}+ 2 \times {\bf 48}^{(+,+)} +3 \times {\bf 84}^{(+,+)}$ & 0.026 & 126.4 & 6.1 \\
(5) &${\bf 7}^{(+,+)}+ {\bf 7}^{(+,-)} +4 \times {\bf 84}^{(+,+)}$ & 0.043 & 126.2 & 3.8 \\
\end{tabular}
\caption{Numerical results of the potential analysis. 
The fermion content, the VEV at the potential minimum, Higgs mass 
and the compactification scale of the fifth dimension are listed. 
The signatures in the superscript of the fields denote $(\eta_R, \eta'_R)$.}
\label{result}
\end{center}
\end{table}
In the obtained results, the case (3) is particularly interesting 
since the compactification is largest among them, whose potential is plotted in Fig \ref{case3}. 

 \section{RGE effect on the weak mixing angle between the weak scale and the compactification scale}
In this section, we study how our prediction for the weak mixing angle is good 
by one-loop renormalization group equation (RGE).  
One-loop RGE for the gauge couplings are  
\begin{align}
\frac{1}{g^2_i(\mu)}-\frac{1}{g^2_i(M_Z)} = \frac{b_i}{8\pi^2} \ln \left( \frac{M_Z}{\mu} \right), 
\end{align}
where the index $i=1,2,3$ labels the 4D gauge coupling $g_{1,2,3}$, 
and $\mu$ is a renormalization scale. 
$b_i$ is the corresponding beta function coefficients given as follows. 
\begin{align}
b_i = -\frac{11}{3} C_2(G) +\frac{2}{3} T_f(R) + \frac{1}{6} T_s(R)
\end{align}
where the first (the second, the last) term is the gauge boson (Weyl fermion, real scalar) contributions, respectively. 
$C_2(G)$ and $T(R)$ are defined in terms of the structure constant and the generators in the representation $R$, 
\begin{align}
C_2(G) \delta^{ab} = f^{abc} f^{dbc}, \qquad T(R) \delta^{ab} = {\rm tr}(T^a(R)T^b(R)). 
\end{align}
For the fundamental representation of $SU(N)$ for example, $C_2({\bf N}) = \frac{N^2-1}{2N}, T({\bf N})=\frac{1}{2}$. 

The RGE between the compactification scale $1/R$ and its half $1/(2R)$, 
and between the half of the compactification scale and the $Z$-boson mass scale $M_Z$ are given as
\begin{align}
\frac{1}{g^2_i(1/R)}-\frac{1}{g^2_i(1/(2R))} &= \frac{1}{8\pi^2} (b_i^{{\rm SM}} + \Delta b_i) \ln \left( \frac{1}{2} \right), \\
\frac{1}{g^2_i(1/(2R))}-\frac{1}{g^2_i(M_Z)} &= \frac{1}{8\pi^2} b_i^{{\rm SM}} \ln \left( 2M_Z R \right), 
\end{align}
where $b_i^{{\rm SM}}$ is the one-loop beta function coefficient of the Standard Model 
and $\Delta b_i$ stands for the beta function coefficients by the contributions from the fields with mass $1/(2R)$. 
Eliminating $g_i^2(1/(2R))$, we obtain 
\begin{align}
\frac{1}{g^2_i(1/R)}-\frac{1}{g^2_i(M_Z)} &= \frac{1}{8\pi^2} (b_i^{{\rm SM}} \ln (M_Z R) - \Delta b_i \ln 2 ). 
\label{RGE1}
\end{align}
The weak mixing angle at the compactification scale can be rewritten into that at the $Z$-boson mass scale 
by using the RGE,  
\begin{align}
\frac{1}{4} = \sin^2 \theta_W  = \frac{g_Y^2}{g_2^2 + g_Y^2} 
= \frac{1}{1 + \frac{3g_2^2}{g_1^2}}
= \left[
1+3\frac{\frac{1}{g_1^2(M_Z)} + \frac{1}{8\pi^2}(b_1^{{\rm SM}} \ln (M_Z R) - \Delta b_1 \ln 2)}
{\frac{1}{g_2^2(M_Z)} + \frac{1}{8\pi^2}(b_2^{{\rm SM}} \ln (M_Z R) - \Delta b_2 \ln 2)}
\right]^{-1},
\end{align}
where a relation $g_Y=g_1/\sqrt{3}$ is put in the third equality and the RGE (\ref{RGE1}) is utilized in the last equality. 
After some algebra, we find an expression for the weak mixing angle at the $Z$-boson masse scale as
\begin{align}
\sin^2 \theta_W(M_Z) 
&= \left[ 4 
-\frac{3 g_2^2(M_z)}{8\pi^2} \left\{ (b_1^{{\rm SM}} - b_2^{{\rm SM}}) \ln (M_Z R)
+ \left( \Delta b_1 -\Delta b_2 \right)\ln 2\right \} \right]^{-1} 
\label{mixingatZ}
\end{align}

In order to obtain the weak mixing angle at the $Z$-boson mass scale, 
we have only to calculate the beta function coefficients $\Delta b_{1,2}$ from the contributions of the fields with mass $1/(2R)$. 
Let us consider each contribution in the case (3) of $3\times {\bf 48}^{(+,+)} + 1 \times {\bf 28}^{(+,+)} + 2\times {\bf 84}^{(+,+)}$ in order. 

As for the gauge field $A_\mu$ and the scalar fields of $A_{5,6}$, 
the representations of the fields with a mass $1/(2R_5)$ are the following. 
\begin{align}
{\bf 48} &\supset 
({\bf 1}, {\bf 2})_{3/2}^{(+, -, +)}
+ ({\bf 1}, \bar{{\bf 2}})_{-3/2}^{(+, -, +)}
+ ({\bf 3}, \bar{{\bf 2}})_{-1/2}^{(-, +, -)}
+ (\bar{{\bf 3}}, {\bf 2})_{1/2}^{(-, +, -)} 
+ ({\bf 1}, {\bf 1})_{-1}^{(+, +, -)}
+ ({\bf 1}, {\bf 1})_{1}^{(+, +, -)}~(A_{\mu}), \\
&
({\bf 1}, {\bf 2})_{3/2}^{(+, +, -)}
+ ({\bf 1}, \bar{{\bf 2}})_{-3/2}^{(+, +, -)}
+ ({\bf 3}, \bar{{\bf 2}})_{-1/2}^{(-, -, +)}
+ (\bar{{\bf 3}}, {\bf 2})_{1/2}^{(-, -, +)}
+ ({\bf 1}, {\bf 1})_{-1}^{(+, -, +)}
+ ({\bf 1}, {\bf 1})_{1}^{(+, -, +)}~(A_{5}), \\
&
({\bf 1}, {\bf 2})_{3/2}^{(-, -, +)}
+ ({\bf 1}, \bar{{\bf 2}})_{-3/2}^{(-, -, +)}
+ ({\bf 3}, \bar{{\bf 2}})_{-1/2}^{(+, +, -)}
+ (\bar{{\bf 3}}, {\bf 2})_{1/2}^{(+, +, -)}
+ ({\bf 1}, {\bf 1})_{-1}^{(-, +, -)}
+ ({\bf 1}, {\bf 1})_{1}^{(-, +, -)}~(A_{6}). 
\end{align}
Note that the fields with $SU(2)_L$ singlet and vanishing hypercharge are neglected 
since they have no contributions to the running of the $SU(2)_L$ and $U(1)_Y$ gauge couplings.  
\noindent
Corresponding beta function coefficients of $A_\mu, A_{5,6}$ are calculated as follows. 
\begin{align}
\Delta b_1({\rm gauge}) 
&= -\left( \frac{11}{3} - \frac{1}{3} \right) \times \left[ \left( \frac{3}{2} \right)^2 \times 4  
+ \left( \frac{1}{2} \right)^2 \times 12 + 1^2 \times 2 \right]  = -\frac{140}{3}, 
\label{betagauge1} \\
\Delta b_2({\rm gauge})  
&= - \frac{11}{3} \times \frac{3}{4} \times 8 + \frac{1}{6} \times 2 \times \frac{1}{2} \times 8 = -\frac{62}{3}, 
\label{betagauge2}
\end{align}
which results in $\Delta b_1-\Delta b_2 = -\frac{78}{3}$. 

As for the fermion in the ${\bf 48}$ dimensional representation, 
Corresponding beta function coefficients are calculated by taking into account Dirac fermions in this calculation. 
\begin{align}
\Delta b_1({\bf 48} \times 3) 
&= \frac{4}{3}  \times \left[ \left( \frac{3}{2} \right)^2 \times 2 \times 2  
+ \left( \frac{1}{2} \right)^2 \times 6 \times 2 + 1^2 \times 2 \right] \times 3 = 56, 
\label{beta481}\\
\Delta b_2({\bf 48} \times 3)  
&=  \frac{4}{3} \times \frac{1}{2} \times 8 \times 3 = 16. 
\label{beta482}
\end{align}
which results in $\Delta b_1-\Delta b_2 = 40$. 

As for the fermion in the ${\bf 28}, {\bf 84}$ dimensional representations, 
the fields of mass $1/(2R_5)$ with $Z_2$ parities $(P_5, P'_5)=(+,-), (-,+)$ in $SU(2)_L \times U(1)_Y$ representation are found as 
\begin{align}
{\bf 28} &\supset ({\bf 3}, {\bf 2})_{1/2}^{(-,+,-)} 
+ ({\bf 1}, {\bf 2})_{-1/2}^{(+,-,+)} + ({\bf 1}, {\bf 1})_{-1}^{(+,+,-)}, 
\nonumber \\
{\bf 84} &\supset 
({\bf 6}, {\bf 2})_{1/2}^{(-,+,-)} 
+ ({\bf 3}, {\bf 2})_{-1/2}^{(+,-,+)}
+ ({\bf 3}, {\bf 1})_{-1}^{(+,+,-)} + ({\bf 1}, {\bf 4})_{3/2}^{(-,+,-)} + ({\bf 1}, {\bf 3})_{1}^{(-,-,+)} \notag \\
&+ ({\bf 1}, {\bf 2})_{1/2}^{(-,+,-)} + ({\bf 1}, {\bf 2})_{-3/2}^{(-,+,-)} 
+ ({\bf 1}, {\bf 1})_{-2}^{(-,-,+)}. 
\end{align}
Corresponding beta function coefficients are calculated as follows. 
\begin{align}
\Delta b_1({\bf 28}) 
&= \frac{4}{3}  \times \left[  \left( \frac{1}{2} \right)^2 \times (6+2) + 1^2  \right]  = 4, 
\label{beta281}\\
\Delta b_2({\bf 28})  
&=  \frac{4}{3} \times \frac{1}{2} \times (3+1) = \frac{8}{3}. 
\label{beta282}
\end{align}
which results in $\Delta b_1-\Delta b_2 = \frac{4}{3}$.
\begin{align}
\Delta b_1({\bf 84} \times 2) 
&= \frac{4}{3}  \times \left[ \left( \frac{3}{2} \right)^2 \times 4 + \left( \frac{1}{2} \right)^2 \times (12 +2)
+ 1^2 \times 3 + (-1)^2 \times 3
\right. \nonumber \\
& \left. 
+ \left( -\frac{1}{2} \right)^2 \times 6 
+ (-2)^2 \times 1 + \left(-\frac{3}{2} \right)^2 \times 2 \right] \times 2 = 76, 
\label{beta841}\\
\Delta b_2({\bf 84} \times 2)  
&=  \frac{4}{3} \times \left[ 5 + \frac{1}{2} \times (6 + 3 + 2) + 2  \right] \times 2 = \frac{100}{3}, 
\label{beta842}
\end{align}
where $T({\bf 2}) = 1/2, T({\bf 3})=2, T({\bf 4})=\frac{(2+2)(2+3)}{4}=5$ for $SU(2)$ are used. 
Therefore, we obtain $\Delta b_1-\Delta b_2 = \frac{128}{3}$.

Collecting all of the results (\ref{betagauge1})-
(\ref{beta482}), (\ref{beta281})-
(\ref{beta842}) together, 
we obtain the beta function coefficients contributing from the fields with mass $1/(2R)$
\begin{align}
\Delta b_1 - \Delta b_2 = -\frac{78}{3} + 40 + \frac{4}{3} + \frac{128}{3} = 58. 
\end{align}

Substituting these results, $\frac{4\pi}{g_2^2(M_Z)} = 29.57 \dots$, $1/R_5 = 7.5$ TeV 
and $b_1^{{\rm SM}}=\frac{41}{6}, b_2^{{\rm SM}}=-\frac{19}{6}$ 
into (\ref{mixingatZ}), 
we find the weak mixing angle at the $Z$-boson mass scale 
\begin{align}
\sin^2 \theta_W(M_Z) 
= \left[ 4 
-\frac{3 g_2^2(M_z)}{8\pi^2} \left\{ (b_1^{{\rm SM}} - b_2^{{\rm SM}}) \ln (M_Z R_5)
+ 58 \ln 2\right \} \right]^{-1} 
\sim 0.246. 
\end{align}
The result is good agreement with the experimental data $\sin^2 \theta_W \sim 0.23$ at the weak scale.

\section{Conclusions}
In this paper, we have proposed a model of 6D $SU(7)$ grand Gauge-Higgs Unification 
with the compactification $S^1/Z_2 \times S^1/Z_2$, 
where the SM Higgs field is embedded in a zero mode of the fifth component of the gauge field $A_5$ 
and which is an extension of 5D $SU(4)$ GHU with a weak mixing angle $\sin^2 \theta_W =1/4$ at the compactification scale. 
As for the SM fermions, the charged leptons can be embedded into a 6D fermion in a two-rank anti-symmetric tensor representation 
and their Yukawa couplings are generated through the higher dimensional gauge coupling. 
The quarks are introduced as 4D fields localized at the fixed points 
since the hypercharges of the bulk field are fixed to be integers or half-integers in our model 
and the quarks with fractional hypercharges cannot be assigned to the bulk fields. 
Yukawa couplings of quarks are localized on the fixed point and the Higgs field is replaced by $A_5$ zero mode. 

We calculate one-loop effective potential of the Higgs field and 
investigate whether the correct pattern of electroweak symmetry breaking 
and a realistic Higgs mass are realized in our model. 
In a case of the gauge field only, it was found that the desired electroweak symmetry breaking did not take place. 
Then, a potential analysis by introducing massless 6D fermions in ${\bf 7}, {\bf 28}, {\bf 48}$ and ${\bf 84}$ representations was performed. 
We have examined all possible combinations up to six fermions and found some solutions 
realizing both of the correct pattern of electroweak symmetry breaking and the realistic Higgs mass 125 GeV. 
The compactification scale of the fifth dimension is also predicted 
corresponding to each solutions obtained above, $1/R_5 \sim 7.5$ TeV for instance. 
Furthermore, one-loop RGE effects of the weak mixing angle was calculated for the case (3) in Table 1. 
The weak mixing angle at the weak scale predicted by one-loop RGE calculations was shown 
to be in good agreement with the experimental data. 
 
There are several issues to be explored although our model is good 
as the starting point toward constructing a more realistic model. 
First, since Yukawa couplings for quarks localized at the fixed point and 
those for leptons generated from the gauge coupling in the bulk are not related in our model, 
the predictions specific to GUT cannot be expected in flavor sector, as it stands. 
In order to avoid this situation, one of the ideas is that all SM fermions are considered to be localized at the fixed points 
and Yukawa couplings are generated by coupling the bulk fermions to the localized SM fermions as in \cite{SSS, MYT}.  
Second, the gauge coupling unification is interesting in our model 
in that the electroweak gauge group $SU(4)$ and the strong gauge group $SU(3)$ 
are unified into the gauge group $SU(7)$ in 6D. 
Quantitative analysis by one-loop RGE of the gauge couplings should be performed in future. 
Proton decay analysis is also important. 
Since the mass of $X, Y$ gauge bosons is small in our model, this causes rapid proton decay if they can couple to quarks and leptons. 
How the dangerous proton decay processes can be forbidden, for instance, by some symmetries 
and what kind of decay modes can be predicted in our model are left for our future work. 
 
\newpage
\appendix 

\section{SU(7) generators}
In this section, the $SU(7)$ generators are summarized. 
{\footnotesize

\begin{align}
  &  T^1 = \frac{1}{2}
  \left(
    \begin{array}{ccccccc}
      0 & 1 & 0 & 0 & 0 & 0 & 0 \\
      1 & 0 & 0 & 0 & 0 & 0 & 0 \\
      0 & 0 & 0 & 0 & 0 & 0 & 0 \\
      0 & 0 & 0 & 0 & 0 & 0 & 0 \\
      0 & 0 & 0 & 0 & 0 & 0 & 0 \\
      0 & 0 & 0 & 0 & 0 & 0 & 0 \\
      0 & 0 & 0 & 0 & 0 & 0 & 0
    \end{array}
  \right),
  &  \quad \qquad
  &  T^2 = \frac{1}{2}
  \left(
    \begin{array}{ccccccc}
      0 & -i & 0 & 0 & 0 & 0 & 0 \\
      i & 0 & 0 & 0 & 0 & 0 & 0 \\
      0 & 0 & 0 & 0 & 0 & 0 & 0 \\
      0 & 0 & 0 & 0 & 0 & 0 & 0 \\
      0 & 0 & 0 & 0 & 0 & 0 & 0 \\
      0 & 0 & 0 & 0 & 0 & 0 & 0 \\
      0 & 0 & 0 & 0 & 0 & 0 & 0
    \end{array}
  \right), \\
  &  T^3 = \frac{1}{2}
  \left(
    \begin{array}{ccccccc}
      1 & 0 & 0 & 0 & 0 & 0 & 0 \\
      0 & -1 & 0 & 0 & 0 & 0 & 0 \\
      0 & 0 & 0 & 0 & 0 & 0 & 0 \\
      0 & 0 & 0 & 0 & 0 & 0 & 0 \\
      0 & 0 & 0 & 0 & 0 & 0 & 0 \\
      0 & 0 & 0 & 0 & 0 & 0 & 0 \\
      0 & 0 & 0 & 0 & 0 & 0 & 0
    \end{array}
  \right),
  &
  &  T^4 = \frac{1}{2}
  \left(
    \begin{array}{ccccccc}
      0 & 0 & 1 & 0 & 0 & 0 & 0 \\
      0 & 0 & 0 & 0 & 0 & 0 & 0 \\
      1 & 0 & 0 & 0 & 0 & 0 & 0 \\
      0 & 0 & 0 & 0 & 0 & 0 & 0 \\
      0 & 0 & 0 & 0 & 0 & 0 & 0 \\
      0 & 0 & 0 & 0 & 0 & 0 & 0 \\
      0 & 0 & 0 & 0 & 0 & 0 & 0
    \end{array}
  \right),  \\
  &  T^5 = \frac{1}{2}
  \left(
    \begin{array}{ccccccc}
      0 & 0 & -i & 0 & 0 & 0 & 0 \\
      0 & 0 & 0 & 0 & 0 & 0 & 0 \\
      i & 0 & 0 & 0 & 0 & 0 & 0 \\
      0 & 0 & 0 & 0 & 0 & 0 & 0 \\
      0 & 0 & 0 & 0 & 0 & 0 & 0 \\
      0 & 0 & 0 & 0 & 0 & 0 & 0 \\
      0 & 0 & 0 & 0 & 0 & 0 & 0
    \end{array}
  \right),
  & \quad \qquad
  &  T^6 = \frac{1}{2}
  \left(
    \begin{array}{ccccccc}
      0 & 0 & 0 & 0 & 0 & 0 & 0 \\
      0 & 0 & 1 & 0 & 0 & 0 & 0 \\
      0 & 1 & 0 & 0 & 0 & 0 & 0 \\
      0 & 0 & 0 & 0 & 0 & 0 & 0 \\
      0 & 0 & 0 & 0 & 0 & 0 & 0 \\
      0 & 0 & 0 & 0 & 0 & 0 & 0 \\
      0 & 0 & 0 & 0 & 0 & 0 & 0
    \end{array}
  \right), \\
  &  T^7 = \frac{1}{2}
  \left(
    \begin{array}{ccccccc}
      0 & 0 & 0 & 0 & 0 & 0 & 0 \\
      0 & 0 & -i & 0 & 0 & 0 & 0 \\
      0 & i & 0 & 0 & 0 & 0 & 0 \\
      0 & 0 & 0 & 0 & 0 & 0 & 0 \\
      0 & 0 & 0 & 0 & 0 & 0 & 0 \\
      0 & 0 & 0 & 0 & 0 & 0 & 0 \\
      0 & 0 & 0 & 0 & 0 & 0 & 0
    \end{array}
  \right),
  &
  &  T^8 = \frac{1}{2\sqrt{3}}
  \left(
    \begin{array}{ccccccc}
      1 & 0 & 0 & 0 & 0 & 0 & 0 \\
      0 & 1 & 0 & 0 & 0 & 0 & 0 \\
      0 & 0 & -2 & 0 & 0 & 0 & 0 \\
      0 & 0 & 0 & 0 & 0 & 0 & 0 \\
      0 & 0 & 0 & 0 & 0 & 0 & 0 \\
      0 & 0 & 0 & 0 & 0 & 0 & 0 \\
      0 & 0 & 0 & 0 & 0 & 0 & 0
    \end{array}
  \right), \\
  &  T^9 = \frac{1}{2}
  \left(
    \begin{array}{ccccccc}
      0 & 0 & 0 & 1 & 0 & 0 & 0 \\
      0 & 0 & 0 & 0 & 0 & 0 & 0 \\
      0 & 0 & 0 & 0 & 0 & 0 & 0 \\
      1 & 0 & 0 & 0 & 0 & 0 & 0 \\
      0 & 0 & 0 & 0 & 0 & 0 & 0 \\
      0 & 0 & 0 & 0 & 0 & 0 & 0 \\
      0 & 0 & 0 & 0 & 0 & 0 & 0
    \end{array}
  \right),
  &
  &  T^{10} = \frac{1}{2}
  \left(
    \begin{array}{ccccccc}
      0 & 0 & 0 & -i & 0 & 0 & 0 \\
      0 & 0 & 0 & 0 & 0 & 0 & 0 \\
      0 & 0 & 0 & 0 & 0 & 0 & 0 \\
      i & 0 & 0 & 0 & 0 & 0 & 0 \\
      0 & 0 & 0 & 0 & 0 & 0 & 0 \\
      0 & 0 & 0 & 0 & 0 & 0 & 0 \\
      0 & 0 & 0 & 0 & 0 & 0 & 0
    \end{array}
  \right), \\
  &  T^{11} = \frac{1}{2}
  \left(
    \begin{array}{ccccccc}
      0 & 0 & 0 & 0 & 0 & 0 & 0 \\
      0 & 0 & 0 & 1 & 0 & 0 & 0 \\
      0 & 0 & 0 & 0 & 0 & 0 & 0 \\
      0 & 1 & 0 & 0 & 0 & 0 & 0 \\
      0 & 0 & 0 & 0 & 0 & 0 & 0 \\
      0 & 0 & 0 & 0 & 0 & 0 & 0 \\
      0 & 0 & 0 & 0 & 0 & 0 & 0
    \end{array}
  \right),
  &
  &  T^{12} = \frac{1}{2}
  \left(
    \begin{array}{ccccccc}
      0 & 0 & 0 & 0 & 0 & 0 & 0 \\
      0 & 0 & 0 & -i & 0 & 0 & 0 \\
      0 & 0 & 0 & 0 & 0 & 0 & 0 \\
      0 & i & 0 & 0 & 0 & 0 & 0 \\
      0 & 0 & 0 & 0 & 0 & 0 & 0 \\
      0 & 0 & 0 & 0 & 0 & 0 & 0 \\
      0 & 0 & 0 & 0 & 0 & 0 & 0
    \end{array}
  \right), 
  \end{align}
  \begin{align}
  &  T^{13} = \frac{1}{2}
  \left(
    \begin{array}{ccccccc}
      0 & 0 & 0 & 0 & 0 & 0 & 0 \\
      0 & 0 & 0 & 0 & 0 & 0 & 0 \\
      0 & 0 & 0 & 1 & 0 & 0 & 0 \\
      0 & 0 & 1 & 0 & 0 & 0 & 0 \\
      0 & 0 & 0 & 0 & 0 & 0 & 0 \\
      0 & 0 & 0 & 0 & 0 & 0 & 0 \\
      0 & 0 & 0 & 0 & 0 & 0 & 0
    \end{array}
  \right),
  &
  &  T^{14} = \frac{1}{2}
  \left(
    \begin{array}{ccccccc}
      0 & 0 & 0 & 0 & 0 & 0 & 0 \\
      0 & 0 & 0 & 0 & 0 & 0 & 0 \\
      0 & 0 & 0 & -i & 0 & 0 & 0 \\
      0 & 0 & i & 0 & 0 & 0 & 0 \\
      0 & 0 & 0 & 0 & 0 & 0 & 0 \\
      0 & 0 & 0 & 0 & 0 & 0 & 0 \\
      0 & 0 & 0 & 0 & 0 & 0 & 0
    \end{array}
  \right), \\
  &  T^{15} = \frac{1}{2}
  \left(
    \begin{array}{ccccccc}
      0 & 0 & 0 & 0 & 0 & 0 & 0 \\
      0 & 0 & 0 & 0 & 0 & 0 & 0 \\
      0 & 0 & 0 & 0 & 0 & 0 & 0 \\
      0 & 0 & 0 & 1 & 0 & 0 & 0 \\
      0 & 0 & 0 & 0 & -1 & 0 & 0 \\
      0 & 0 & 0 & 0 & 0 & 0 & 0 \\
      0 & 0 & 0 & 0 & 0 & 0 & 0
    \end{array}
  \right),
  &
  &  T^{16} = \frac{1}{2}
  \left(
    \begin{array}{ccccccc}
      0 & 0 & 0 & 0 & 1 & 0 & 0 \\
      0 & 0 & 0 & 0 & 0 & 0 & 0 \\
      0 & 0 & 0 & 0 & 0 & 0 & 0 \\
      0 & 0 & 0 & 0 & 0 & 0 & 0 \\
      1 & 0 & 0 & 0 & 0 & 0 & 0 \\
      0 & 0 & 0 & 0 & 0 & 0 & 0 \\
      0 & 0 & 0 & 0 & 0 & 0 & 0
    \end{array}
  \right), \\
  &  T^{17} = \frac{1}{2}
  \left(
    \begin{array}{ccccccc}
      0 & 0 & 0 & 0 & -i & 0 & 0 \\
      0 & 0 & 0 & 0 & 0 & 0 & 0 \\
      0 & 0 & 0 & 0 & 0 & 0 & 0 \\
      0 & 0 & 0 & 0 & 0 & 0 & 0 \\
      i & 0 & 0 & 0 & 0 & 0 & 0 \\
      0 & 0 & 0 & 0 & 0 & 0 & 0 \\
      0 & 0 & 0 & 0 & 0 & 0 & 0
    \end{array}
  \right),
  &
  &  T^{18} = \frac{1}{2}
  \left(
    \begin{array}{ccccccc}
      0 & 0 & 0 & 0 & 0 & 0 & 0 \\
      0 & 0 & 0 & 0 & 1 & 0 & 0 \\
      0 & 0 & 0 & 0 & 0 & 0 & 0 \\
      0 & 0 & 0 & 0 & 0 & 0 & 0 \\
      0 & 1 & 0 & 0 & 0 & 0 & 0 \\
      0 & 0 & 0 & 0 & 0 & 0 & 0 \\
      0 & 0 & 0 & 0 & 0 & 0 & 0
    \end{array}
  \right), \\
  &  T^{19} = \frac{1}{2}
  \left(
    \begin{array}{ccccccc}
      0 & 0 & 0 & 0 & 0 & 0 & 0 \\
      0 & 0 & 0 & 0 & -i & 0 & 0 \\
      0 & 0 & 0 & 0 & 0 & 0 & 0 \\
      0 & 0 & 0 & 0 & 0 & 0 & 0 \\
      0 & i & 0 & 0 & 0 & 0 & 0 \\
      0 & 0 & 0 & 0 & 0 & 0 & 0 \\
      0 & 0 & 0 & 0 & 0 & 0 & 0
    \end{array}
  \right),
  & 
  &  T^{20} = \frac{1}{2}
  \left(
    \begin{array}{ccccccc}
      0 & 0 & 0 & 0 & 0 & 0 & 0 \\
      0 & 0 & 0 & 0 & 0 & 0 & 0 \\
      0 & 0 & 0 & 0 & 1 & 0 & 0 \\
      0 & 0 & 0 & 0 & 0 & 0 & 0 \\
      0 & 0 & 1 & 0 & 0 & 0 & 0 \\
      0 & 0 & 0 & 0 & 0 & 0 & 0 \\
      0 & 0 & 0 & 0 & 0 & 0 & 0
    \end{array}
  \right), \\
  &  T^{21} = \frac{1}{2}
  \left(
    \begin{array}{ccccccc}
      0 & 0 & 0 & 0 & 0 & 0 & 0 \\
      0 & 0 & 0 & 0 & 0 & 0 & 0 \\
      0 & 0 & 0 & 0 & -i & 0 & 0 \\
      0 & 0 & 0 & 0 & 0 & 0 & 0 \\
      0 & 0 & i & 0 & 0 & 0 & 0 \\
      0 & 0 & 0 & 0 & 0 & 0 & 0 \\
      0 & 0 & 0 & 0 & 0 & 0 & 0
    \end{array}
  \right),
  &
  &  T^{22} = \frac{1}{2}
  \left(
    \begin{array}{ccccccc}
      0 & 0 & 0 & 0 & 0 & 0 & 0 \\
      0 & 0 & 0 & 0 & 0 & 0 & 0 \\
      0 & 0 & 0 & 0 & 0 & 0 & 0 \\
      0 & 0 & 0 & 0 & 1 & 0 & 0 \\
      0 & 0 & 0 & 1 & 0 & 0 & 0 \\
      0 & 0 & 0 & 0 & 0 & 0 & 0 \\
      0 & 0 & 0 & 0 & 0 & 0 & 0
    \end{array}
  \right), \notag \\
  &  T^{23} = \frac{1}{2}
  \left(
    \begin{array}{ccccccc}
      0 & 0 & 0 & 0 & 0 & 0 & 0 \\
      0 & 0 & 0 & 0 & 0 & 0 & 0 \\
      0 & 0 & 0 & 0 & 0 & 0 & 0 \\
      0 & 0 & 0 & 0 & -i & 0 & 0 \\
      0 & 0 & 0 & i & 0 & 0 & 0 \\
      0 & 0 & 0 & 0 & 0 & 0 & 0 \\
      0 & 0 & 0 & 0 & 0 & 0 & 0
    \end{array}
  \right),
  & \quad \qquad
  &  T^{24} = \frac{1}{2\sqrt{3}}
  \left(
    \begin{array}{ccccccc}
      0 & 0 & 0 & 0 & 0 & 0 & 0 \\
      0 & 0 & 0 & 0 & 0 & 0 & 0 \\
      0 & 0 & 0 & 0 & 0 & 0 & 0 \\
      0 & 0 & 0 & 1 & 0 & 0 & 0 \\
      0 & 0 & 0 & 0 & 1 & 0 & 0 \\
      0 & 0 & 0 & 0 & 0 & -2 & 0 \\
      0 & 0 & 0 & 0 & 0 & 0 & 0
    \end{array}
  \right), \\
  &  T^{25} = \frac{1}{2}
  \left(
    \begin{array}{ccccccc}
      0 & 0 & 0 & 0 & 0 & 1 & 0 \\
      0 & 0 & 0 & 0 & 0 & 0 & 0 \\
      0 & 0 & 0 & 0 & 0 & 0 & 0 \\
      0 & 0 & 0 & 0 & 0 & 0 & 0 \\
      0 & 0 & 0 & 0 & 0 & 0 & 0 \\
      1 & 0 & 0 & 0 & 0 & 0 & 0 \\
      0 & 0 & 0 & 0 & 0 & 0 & 0
    \end{array}
  \right),
  &
  &  T^{26} = \frac{1}{2}
  \left(
    \begin{array}{ccccccc}
      0 & 0 & 0 & 0 & 0 & -i & 0 \\
      0 & 0 & 0 & 0 & 0 & 0 & 0 \\
      0 & 0 & 0 & 0 & 0 & 0 & 0 \\
      0 & 0 & 0 & 0 & 0 & 0 & 0 \\
      0 & 0 & 0 & 0 & 0 & 0 & 0 \\
      i & 0 & 0 & 0 & 0 & 0 & 0 \\
      0 & 0 & 0 & 0 & 0 & 0 & 0
    \end{array}
  \right), 
   \end{align}
   \begin{align}
  &  T^{27} = \frac{1}{2}
  \left(
    \begin{array}{ccccccc}
      0 & 0 & 0 & 0 & 0 & 0 & 0 \\
      0 & 0 & 0 & 0 & 0 & 1 & 0 \\
      0 & 0 & 0 & 0 & 0 & 0 & 0 \\
      0 & 0 & 0 & 0 & 0 & 0 & 0 \\
      0 & 0 & 0 & 0 & 0 & 0 & 0 \\
      0 & 1 & 0 & 0 & 0 & 0 & 0 \\
      0 & 0 & 0 & 0 & 0 & 0 & 0
    \end{array}
  \right),
  &
  &  T^{28} = \frac{1}{2}
  \left(
    \begin{array}{ccccccc}
      0 & 0 & 0 & 0 & 0 & 0 & 0 \\
      0 & 0 & 0 & 0 & 0 & -i & 0 \\
      0 & 0 & 0 & 0 & 0 & 0 & 0 \\
      0 & 0 & 0 & 0 & 0 & 0 & 0 \\
      0 & 0 & 0 & 0 & 0 & 0 & 0 \\
      0 & i & 0 & 0 & 0 & 0 & 0 \\
      0 & 0 & 0 & 0 & 0 & 0 & 0
    \end{array}
  \right), \\
  &  T^{29} = \frac{1}{2}
  \left(
    \begin{array}{ccccccc}
      0 & 0 & 0 & 0 & 0 & 0 & 0 \\
      0 & 0 & 0 & 0 & 0 & 0 & 0 \\
      0 & 0 & 0 & 0 & 0 & 1 & 0 \\
      0 & 0 & 0 & 0 & 0 & 0 & 0 \\
      0 & 0 & 0 & 0 & 0 & 0 & 0 \\
      0 & 0 & 1 & 0 & 0 & 0 & 0 \\
      0 & 0 & 0 & 0 & 0 & 0 & 0
    \end{array}
  \right),
  &
  &  T^{30} = \frac{1}{2}
  \left(
    \begin{array}{ccccccc}
      0 & 0 & 0 & 0 & 0 & 0 & 0 \\
      0 & 0 & 0 & 0 & 0 & 0 & 0 \\
      0 & 0 & 0 & 0 & 0 & -i & 0 \\
      0 & 0 & 0 & 0 & 0 & 0 & 0 \\
      0 & 0 & 0 & 0 & 0 & 0 & 0 \\
      0 & 0 & i & 0 & 0 & 0 & 0 \\
      0 & 0 & 0 & 0 & 0 & 0 & 0
    \end{array}
  \right), \\
  &  T^{31} = \frac{1}{2}
  \left(
    \begin{array}{ccccccc}
      0 & 0 & 0 & 0 & 0 & 0 & 0 \\
      0 & 0 & 0 & 0 & 0 & 0 & 0 \\
      0 & 0 & 0 & 0 & 0 & 0 & 0 \\
      0 & 0 & 0 & 0 & 0 & 1 & 0 \\
      0 & 0 & 0 & 0 & 0 & 0 & 0 \\
      0 & 0 & 0 & 1 & 0 & 0 & 0 \\
      0 & 0 & 0 & 0 & 0 & 0 & 0
    \end{array}
  \right),
  &
  &  T^{32} = \frac{1}{2}
  \left(
    \begin{array}{ccccccc}
      0 & 0 & 0 & 0 & 0 & 0 & 0 \\
      0 & 0 & 0 & 0 & 0 & 0 & 0 \\
      0 & 0 & 0 & 0 & 0 & 0 & 0 \\
      0 & 0 & 0 & 0 & 0 & -i & 0 \\
      0 & 0 & 0 & 0 & 0 & 0 & 0 \\
      0 & 0 & 0 & i & 0 & 0 & 0 \\
      0 & 0 & 0 & 0 & 0 & 0 & 0
    \end{array}
  \right), \\
  &  T^{33} = \frac{1}{2}
  \left(
    \begin{array}{ccccccc}
      0 & 0 & 0 & 0 & 0 & 0 & 0 \\
      0 & 0 & 0 & 0 & 0 & 0 & 0 \\
      0 & 0 & 0 & 0 & 0 & 0 & 0 \\
      0 & 0 & 0 & 0 & 0 & 0 & 0 \\
      0 & 0 & 0 & 0 & 0 & 1 & 0 \\
      0 & 0 & 0 & 0 & 1 & 0 & 0 \\
      0 & 0 & 0 & 0 & 0 & 0 & 0
    \end{array}
  \right),
  &
  &  T^{34} = \frac{1}{2}
  \left(
    \begin{array}{ccccccc}
      0 & 0 & 0 & 0 & 0 & 0 & 0 \\
      0 & 0 & 0 & 0 & 0 & 0 & 0 \\
      0 & 0 & 0 & 0 & 0 & 0 & 0 \\
      0 & 0 & 0 & 0 & 0 & 0 & 0 \\
      0 & 0 & 0 & 0 & 0 & -i & 0 \\
      0 & 0 & 0 & 0 & i & 0 & 0 \\
      0 & 0 & 0 & 0 & 0 & 0 & 0
    \end{array}
  \right), \\
  &  T^{35} = \frac{1}{2\sqrt{6}}
  \left(
    \begin{array}{ccccccc}
      0 & 0 & 0 & 0 & 0 & 0 & 0 \\
      0 & 0 & 0 & 0 & 0 & 0 & 0 \\
      0 & 0 & 0 & 0 & 0 & 0 & 0 \\
      0 & 0 & 0 & 1 & 0 & 0 & 0 \\
      0 & 0 & 0 & 0 & 1 & 0 & 0 \\
      0 & 0 & 0 & 0 & 0 & 1 & 0 \\
      0 & 0 & 0 & 0 & 0 & 0 & -3
    \end{array}
  \right),
  &
  &  T^{36} = \frac{1}{2}
  \left(
    \begin{array}{ccccccc}
      0 & 0 & 0 & 0 & 0 & 0 & 1 \\
      0 & 0 & 0 & 0 & 0 & 0 & 0 \\
      0 & 0 & 0 & 0 & 0 & 0 & 0 \\
      0 & 0 & 0 & 0 & 0 & 0 & 0 \\
      0 & 0 & 0 & 0 & 0 & 0 & 0 \\
      0 & 0 & 0 & 0 & 0 & 0 & 0 \\
      1 & 0 & 0 & 0 & 0 & 0 & 0
    \end{array}
  \right),\\
  &  T^{37} = \frac{1}{2}
  \left(
    \begin{array}{ccccccc}
      0 & 0 & 0 & 0 & 0 & 0 & -i \\
      0 & 0 & 0 & 0 & 0 & 0 & 0 \\
      0 & 0 & 0 & 0 & 0 & 0 & 0 \\
      0 & 0 & 0 & 0 & 0 & 0 & 0 \\
      0 & 0 & 0 & 0 & 0 & 0 & 0 \\
      0 & 0 & 0 & 0 & 0 & 0 & 0 \\
      i & 0 & 0 & 0 & 0 & 0 & 0
    \end{array}
  \right),
  &  
  &  T^{38} = \frac{1}{2}
  \left(
    \begin{array}{ccccccc}
      0 & 0 & 0 & 0 & 0 & 0 & 0 \\
      0 & 0 & 0 & 0 & 0 & 0 & 1 \\
      0 & 0 & 0 & 0 & 0 & 0 & 0 \\
      0 & 0 & 0 & 0 & 0 & 0 & 0 \\
      0 & 0 & 0 & 0 & 0 & 0 & 0 \\
      0 & 0 & 0 & 0 & 0 & 0 & 0 \\
      0 & 1 & 0 & 0 & 0 & 0 & 0
    \end{array}
  \right), \\
  &  T^{39} = \frac{1}{2}
  \left(
    \begin{array}{ccccccc}
      0 & 0 & 0 & 0 & 0 & 0 & 0 \\
      0 & 0 & 0 & 0 & 0 & 0 & -i \\
      0 & 0 & 0 & 0 & 0 & 0 & 0 \\
      0 & 0 & 0 & 0 & 0 & 0 & 0 \\
      0 & 0 & 0 & 0 & 0 & 0 & 0 \\
      0 & 0 & 0 & 0 & 0 & 0 & 0 \\
      0 & i & 0 & 0 & 0 & 0 & 0
    \end{array}
  \right),
  &
  &  T^{40} = \frac{1}{2}
  \left(
    \begin{array}{ccccccc}
      0 & 0 & 0 & 0 & 0 & 0 & 0 \\
      0 & 0 & 0 & 0 & 0 & 0 & 0 \\
      0 & 0 & 0 & 0 & 0 & 0 & 1 \\
      0 & 0 & 0 & 0 & 0 & 0 & 0 \\
      0 & 0 & 0 & 0 & 0 & 0 & 0 \\
      0 & 0 & 0 & 0 & 0 & 0 & 0 \\
      0 & 0 & 1 & 0 & 0 & 0 & 0
    \end{array}
  \right),  
\end{align}
\begin{align}
  &  T^{41} = \frac{1}{2}
  \left(
    \begin{array}{ccccccc}
      0 & 0 & 0 & 0 & 0 & 0 & 0 \\
      0 & 0 & 0 & 0 & 0 & 0 & 0 \\
      0 & 0 & 0 & 0 & 0 & 0 & -i \\
      0 & 0 & 0 & 0 & 0 & 0 & 0 \\
      0 & 0 & 0 & 0 & 0 & 0 & 0 \\
      0 & 0 & 0 & 0 & 0 & 0 & 0 \\
      0 & 0 & i & 0 & 0 & 0 & 0
    \end{array}
  \right),
  & \quad \qquad
  &  T^{42} = \frac{1}{2}
  \left(
    \begin{array}{ccccccc}
      0 & 0 & 0 & 0 & 0 & 0 & 0 \\
      0 & 0 & 0 & 0 & 0 & 0 & 0 \\
      0 & 0 & 0 & 0 & 0 & 0 & 0 \\
      0 & 0 & 0 & 0 & 0 & 0 & 1 \\
      0 & 0 & 0 & 0 & 0 & 0 & 0 \\
      0 & 0 & 0 & 0 & 0 & 0 & 0 \\
      0 & 0 & 0 & 1 & 0 & 0 & 0
    \end{array}
  \right), \\
  &  T^{43} = \frac{1}{2}
  \left(
    \begin{array}{ccccccc}
      0 & 0 & 0 & 0 & 0 & 0 & 0 \\
      0 & 0 & 0 & 0 & 0 & 0 & 0 \\
      0 & 0 & 0 & 0 & 0 & 0 & 0 \\
      0 & 0 & 0 & 0 & 0 & 0 & -i \\
      0 & 0 & 0 & 0 & 0 & 0 & 0 \\
      0 & 0 & 0 & 0 & 0 & 0 & 0 \\
      0 & 0 & 0 & i & 0 & 0 & 0
    \end{array}
  \right),
  &
  &  T^{44} = \frac{1}{2}
  \left(
    \begin{array}{ccccccc}
      0 & 0 & 0 & 0 & 0 & 0 & 0 \\
      0 & 0 & 0 & 0 & 0 & 0 & 0 \\
      0 & 0 & 0 & 0 & 0 & 0 & 0 \\
      0 & 0 & 0 & 0 & 0 & 0 & 0 \\
      0 & 0 & 0 & 0 & 0 & 0 & 1 \\
      0 & 0 & 0 & 0 & 0 & 0 & 0 \\
      0 & 0 & 0 & 0 & 1 & 0 & 0
    \end{array}
  \right), \\
  &  T^{45} = \frac{1}{2}
  \left(
    \begin{array}{ccccccc}
      0 & 0 & 0 & 0 & 0 & 0 & 0 \\
      0 & 0 & 0 & 0 & 0 & 0 & 0 \\
      0 & 0 & 0 & 0 & 0 & 0 & 0 \\
      0 & 0 & 0 & 0 & 0 & 0 & 0 \\
      0 & 0 & 0 & 0 & 0 & 0 & -i \\
      0 & 0 & 0 & 0 & 0 & 0 & 0 \\
      0 & 0 & 0 & 0 & i & 0 & 0
    \end{array}
  \right),
  &
  &  T^{46} = \frac{1}{2}
  \left(
    \begin{array}{ccccccc}
      0 & 0 & 0 & 0 & 0 & 0 & 0 \\
      0 & 0 & 0 & 0 & 0 & 0 & 0 \\
      0 & 0 & 0 & 0 & 0 & 0 & 0 \\
      0 & 0 & 0 & 0 & 0 & 0 & 0 \\
      0 & 0 & 0 & 0 & 0 & 0 & 0 \\
      0 & 0 & 0 & 0 & 0 & 0 & 1 \\
      0 & 0 & 0 & 0 & 0 & 1 & 0
    \end{array}
  \right), \\
  &  T^{47} = \frac{1}{2}
  \left(
    \begin{array}{ccccccc}
      0 & 0 & 0 & 0 & 0 & 0 & 0 \\
      0 & 0 & 0 & 0 & 0 & 0 & 0 \\
      0 & 0 & 0 & 0 & 0 & 0 & 0 \\
      0 & 0 & 0 & 0 & 0 & 0 & 0 \\
      0 & 0 & 0 & 0 & 0 & 0 & 0 \\
      0 & 0 & 0 & 0 & 0 & 0 & -i \\
      0 & 0 & 0 & 0 & 0 & i & 0
    \end{array}
  \right),
  &
  &  T^{48} = \frac{1}{2\sqrt{42}}
  \left(
    \begin{array}{ccccccc}
      4 & 0 & 0 & 0 & 0 & 0 & 0 \\
      0 & 4 & 0 & 0 & 0 & 0 & 0 \\
      0 & 0 & 4 & 0 & 0 & 0 & 0 \\
      0 & 0 & 0 & -3 & 0 & 0 & 0 \\
      0 & 0 & 0 & 0 & -3 & 0 & 0 \\
      0 & 0 & 0 & 0 & 0 & -3 & 0 \\
      0 & 0 & 0 & 0 & 0 & 0 & -3
    \end{array}
  \right).
\end{align}
}


\section{Calculation of KK masses for the 6D gauge bosons}
\subsection{Diagonalization of the mass matrices for 4D gauge field}
\label{subsec:gaugeKKmass}
In this subsection, we calculate the KK masses for the 4D gauge boson. 
The relevant mass terms is given by 
\begin{align*}
  &\frac{1}{2} A_\mu^a(D_5 D^5 A^\mu)^a
  +\frac{1}{2} A_\mu^a(D_6 D^6 A^\mu)^a \\
  &\qquad
  = - \frac{1}{2} A_\mu^a 
  \left\{
  \left(I^{ab}\partial_5 + f^{a\,44\,b} \frac{\alpha}{R_5}\right)
  \left(I^{bc}\partial_5 + f^{b\,44\,c} \frac{\alpha}{R_5}\right)
  + I^{ac}\partial_6 \partial_6 \right\}
  A^{\mu \, c} + \cdots, 
  \stepcounter{equation}\tag{\theequation}
\end{align*}
where $\ev{A_5^{44}} = \frac{\alpha}{g R_5}$. 
Nonvanishing $SU(7)$ structure constants $f^{abc}$ including 44 index are
\begin{align}
  &f^{15 \, 44 \, 45} = -\frac{1}{2}, \quad
  f^{16 \, 44 \, 37} = \frac{1}{2}, \quad
  f^{17 \, 44 \, 36} = -\frac{1}{2}, \quad
  f^{18 \, 44 \, 39} = \frac{1}{2}, 
  f^{19 \, 44 \, 38} = -\frac{1}{2}, \notag \\
  &f^{20 \, 44 \, 41} = \frac{1}{2}, \quad
  f^{21 \, 44 \, 40} = -\frac{1}{2}, \quad
  f^{22 \, 44 \, 43} = \frac{1}{2}, \quad
  f^{23 \, 44 \, 42} = -\frac{1}{2}, \quad
  f^{24 \, 44 \, 45} = \frac{1}{2\sqrt{3}}, \notag \\
 & f^{33 \, 44 \, 47} = \frac{1}{2}, \quad
  f^{34 \, 44 \, 46} = \frac{1}{2}, \quad
  f^{35 \, 44 \, 45} = \sqrt{\frac{2}{3}}. 
\end{align}
Let us diagonalize the mass matrices in order. 

First, we focus on the KK masses matrix in $a=15,24,35,45$ sector. 
\begin{align*}
  &
  \frac{1}{2} \int dx_5 \int dx_6 \ 
  \left(
  \begin{array}{cccc}
    A_{\mu\ppp}^{15} & A_{\mu\ppp}^{24} & A_{\mu\ppp}^{35} & A_{\mu\pmm}^{45}
  \end{array}
  \right) \\
  & \quad \times
  \left(
  \begin{array}{cccc}
    -\partial_5^2 - \partial_6^2 + \frac{\alpha^2}{4 R_5^2} 
    & -\frac{\alpha^2}{4\sqrt{3}R_5^2}
    & -\frac{\alpha^2}{\sqrt{6} R_5^2}
    &\frac{\alpha}{R_5} \partial_5 \\
    -\frac{\alpha^2}{4\sqrt{3} R_5^2}
    & -\partial_5^2 - \partial_6^2 + \frac{\alpha^2}{12 R_5^2}
    & \frac{\alpha^2}{3\sqrt{2} R_5^2}
    & -\frac{\alpha}{\sqrt{3} R_5} \partial_5 \\
    -\frac{\alpha^2}{\sqrt{6} R_5^2}
    & \frac{\alpha^2}{3\sqrt{2} R_5^2}
    & -\partial_5^2 - \partial_6^2 + \frac{2 \alpha^2}{3 R_5^2}
    & -2\sqrt{\frac{2}{3}}\frac{\alpha}{R_5} \partial_5 \\
    -\frac{\alpha}{R_5} \partial_5
    & \frac{\alpha}{\sqrt{3} R_5} \partial_5
    & 2\sqrt{\frac{2}{3}}\frac{\alpha}{R_5} \partial_5
    & -\partial_5^2 - \partial_6^2 + \frac{\alpha^2}{R_5^2}
  \end{array}
  \right)
  \left(
  \begin{array}{c}
    A_{\ppp}^{\mu \, 15} \\ A_{\ppp}^{\mu \, 24} \\ A_{\ppp}^{\mu \, 35} \\ A_{\pmm}^{\mu \, 45}
  \end{array}
  \right)\\
  &=
  \frac{1}{2}
  \left(
  \begin{array}{ccc}
    A_\mu^{15(0,0)} & A_\mu^{24(0,0)} & A_\mu^{35(0,0)}
  \end{array}
  \right) \\
  & \quad \times
  \left(
  \begin{array}{ccc}
    \frac{\alpha^2}{4 R_5^2} & -\frac{\alpha^2}{4\sqrt{3} R_5^2}
    & -\frac{\alpha^2}{\sqrt{6} R_5^2} \\
    -\frac{\alpha^2}{4\sqrt{3} R_5^2} & \frac{\alpha^2}{12 R_5^2}
    & \frac{\alpha^2}{3\sqrt{2}R_5^2} \\
    -\frac{\alpha^2}{\sqrt{6} R_5^2} & \frac{\alpha^2}{3\sqrt{2}R_5^2}
    & \frac{2 \alpha^2}{3 R_5^2} 
  \end{array}
  \right)
  \left(
  \begin{array}{c}
    A^{\mu \, 15(0,0)} \\ A^{\mu \, 24(0,0)} \\ A^{\mu \, 35(0,0)}
  \end{array}
  \right)\\
  &+
  \frac{1}{2}
  \sum_{n=1}^\infty
  \left(
  \begin{array}{cccc}
    A_\mu^{15(n,0)} & A_\mu^{24(n,0)} & A_\mu^{35(n,0)} & A_\mu^{45(n,0)}
  \end{array}
  \right) \\
  & \quad \times
  \left(
  \begin{array}{cccc}
    \frac{n^2}{R_5^2} + \frac{\alpha^2}{4 R_5^2} & -\frac{\alpha^2}{4\sqrt{3} R_5^2}
    & -\frac{\alpha^2}{\sqrt{6} R_5^2} & \frac{n \alpha}{R_5^2} \\
    -\frac{\alpha^2}{4\sqrt{3} R_5^2} & \frac{n^2}{R_5^2} + \frac{\alpha^2}{12 R_5^2}
    & \frac{\alpha^2}{3\sqrt{2}R_5^2} & -\frac{n \alpha}{\sqrt{3} R_5^2} \\
    -\frac{\alpha^2}{\sqrt{6} R_5^2} & \frac{\alpha^2}{3\sqrt{2}R_5^2}
    & \frac{n^2}{R_5^2} + \frac{2 \alpha^2}{3 R_5^2} & -2\sqrt{\frac{2}{3}}\frac{n \alpha}{R_5^2} \\
    \frac{n \alpha}{R_5^2} & -\frac{n \alpha}{\sqrt{3} R_5^2} & -2\sqrt{\frac{2}{3}}\frac{n \alpha}{R_5^2} 
    & \frac{n^2}{R_5^2} + \frac{\alpha^2}{R_5^2}
  \end{array}
  \right)
  \left(
  \begin{array}{c}
    A^{\mu \, 15(n,0)} \\ A^{\mu \, 24(n,0)} \\ A^{\mu \, 35(n,0)} \\ A^{\mu \, 45(n,0)}
  \end{array}
  \right)\\
  &+
  \frac{1}{2}
  \sum_{m=1}^\infty
  \left(
  \begin{array}{ccc}
    A_\mu^{15(0,m)} & A_\mu^{24(0,m)} & A_\mu^{35(0,m)}
  \end{array}
  \right) \\
  & \quad \times
  \left(
  \begin{array}{ccc}
    \frac{m^2}{R_6^2} + \frac{\alpha^2}{4 R_5^2} & -\frac{\alpha^2}{4\sqrt{3} R_5^2}
    & -\frac{\alpha^2}{\sqrt{6} R_5^2} \\
    -\frac{\alpha^2}{4\sqrt{3} R_5^2} & \frac{m^2}{R_6^2} + \frac{\alpha^2}{12 R_5^2}
    & \frac{\alpha^2}{3\sqrt{2}R_5^2} \\
    -\frac{\alpha^2}{\sqrt{6} R_5^2} & \frac{\alpha^2}{3\sqrt{2}R_5^2}
    & \frac{m^2}{R_6^2} + \frac{2 \alpha^2}{3 R_5^2}
  \end{array}
  \right)
  \left(
  \begin{array}{c}
    A^{\mu \, 15(0,m)} \\ A^{\mu \, 24(0,m)} \\ A^{\mu \, 35(0,m)}
  \end{array}
  \right) \\
  &+
  \frac{1}{2}
  \sum_{n=1}^\infty \sum_{m=1}^\infty 
  \left(
  \begin{array}{cccc}
    A_\mu^{15(n,m)} & A_\mu^{24(n,m)} & A_\mu^{35(n,m)} & A_\mu^{45(n,m)}
  \end{array}
  \right) \\
  & \quad \times
  \left(
  \begin{array}{cccc}
    \frac{n^2}{R_5^2} + \frac{m^2}{R_6^2} + \frac{\alpha^2}{4 R_5^2} & -\frac{\alpha^2}{4\sqrt{3} R_5^2}
    & -\frac{\alpha^2}{\sqrt{6} R_5^2} & \frac{n \alpha}{R_5^2} \\
    -\frac{\alpha^2}{4\sqrt{3} R_5^2} & \frac{n^2}{R_5^2} + \frac{m^2}{R_6^2} + \frac{\alpha^2}{12 R_5^2}
    & \frac{\alpha^2}{3\sqrt{2}R_5^2} & -\frac{n \alpha}{\sqrt{3} R_5^2} \\
    -\frac{\alpha^2}{\sqrt{6} R_5^2} & \frac{\alpha^2}{3\sqrt{2}R_5^2}
    & \frac{n^2}{R_5^2} + \frac{m^2}{R_6^2} + \frac{2 \alpha^2}{3 R_5^2} & -2\sqrt{\frac{2}{3}}\frac{n \alpha}{R_5^2} \\
    \frac{n \alpha}{R_5^2} & -\frac{n \alpha}{\sqrt{3} R_5^2} & -2\sqrt{\frac{2}{3}}\frac{n \alpha}{R_5^2} 
    & \frac{n^2}{R_5^2} + \frac{m^2}{R_6^2} + \frac{\alpha^2}{R_5^2}
  \end{array}
  \right)
  \left(
  \begin{array}{c}
    A^{\mu \, 15(n,m)} \\ A^{\mu \, 24(n,m)} \\ A^{\mu \, 35(n,m)} \\ A^{\mu \, 45(n,m)}
  \end{array}
  \right)
  \stepcounter{equation}\tag{\theequation}
\end{align*}
Diagonalizing these mass matrices, we obtain the KK mass eigenvalues. 
\begin{align}
  m_{n,m}^2 =&
  0,\ 0,\ 
  \frac{\alpha^2}{R_5^2}, \
  \frac{n^2}{R_5^2}, \
  \frac{n^2}{R_5^2}, \
  \frac{(n\pm\alpha)^2}{R_5^2}, \notag \\
  &\frac{m^2}{R_6^2}, \
  \frac{m^2}{R_6^2}, \
  \frac{\alpha^2}{R_5^2}+\frac{m^2}{R_6^2}, \
  \frac{n^2}{R_5^2}+\frac{m^2}{R_6^2}, \
  \frac{n^2}{R_5^2}+\frac{m^2}{R_6^2}, \
  \frac{(n\pm\alpha)^2}{R_5^2}+\frac{m^2}{R_6^2}, 
\end{align}
where $m, n=1, \cdots, \infty$. 

The mass matrix in $a=16-37$ sector is
\begin{align*}
  &\frac{1}{2} \int dx_5 \int dx_6 \ 
  \left(
  \begin{array}{cc}
    A_{\mu\mpm}^{16} & A_{\mu\mmp}^{37}
  \end{array}
  \right)
  \left(
  \begin{array}{cc}
    -\partial_5^2 -\partial_6^2 + \frac{\alpha^2}{4 R_5^2}
    & -\frac{\alpha}{R_5} \partial_5 \\
    \frac{\alpha}{R_5} \partial_5
    & -\partial_5^2 -\partial_6^2 + \frac{\alpha^2}{4 R_5^2}
  \end{array}
  \right)
  \left(
  \begin{array}{c}
    A_{\mpm}^{\mu \, 16} \\ A_{\mmp}^{\mu \, 37}
  \end{array}
  \right)\\
  &=
  \frac{1}{2}
  \sum_{n=0}^\infty \sum_{m=1}^\infty 
  \left(
  \begin{array}{cc}
    A_\mu^{16(n,m)} & A_\mu^{37(n,m)}
  \end{array}
  \right)
  \left(
  \begin{array}{cc}
    \frac{(n+\frac{1}{2})^2}{R_5^2} + \frac{m^2}{R_6^2} + \frac{\alpha^2}{4 R_5^2} & - \frac{(n+\frac{1}{2}) \alpha}{R_5^2} \\
    -\frac{(n+\frac{1}{2}) \alpha}{R_5^2} & \frac{(n+\frac{1}{2})^2}{R_5^2} + \frac{m^2}{R_6^2} + \frac{\alpha^2}{4 R_5^2}
  \end{array}
  \right)
  \left(
  \begin{array}{c}
    A^{\mu \, 16(n,m)} \\ A^{\mu \, 37(n,m)}
  \end{array}
  \right)
  \stepcounter{equation}\tag{\theequation}
\end{align*}
Diagonalizing these mass matrices, we obtain the KK mass eigenvalues. 
\begin{align}
  m_{n,m}^2 =
  \frac{(n+\frac{1}{2}\pm\frac{\alpha}{2})^2}{R_5^2}
  +\frac{m^2}{R_6^2}, 
\end{align}
where $n=0, \cdots, \infty, m=1, \cdots, \infty$. 
The KK mass eigenvalues in $a=18-39, 20-41$ sectors can be similarly obtained.

The mass matrix $a=17-36$ sector is 
\begin{align*}
  &\frac{1}{2} \int dx_5 \int dx_6 \ 
  \left(
  \begin{array}{cc}
    A_{\mu\mpm}^{17} & A_{\mu\mmp}^{36}
  \end{array}
  \right)
  \left(
  \begin{array}{cc}
    -\partial_5^2 - \partial_6^2 + \frac{\alpha^2}{4 R_5^2}
    & \frac{\alpha}{R_5} \partial_5 \\
    -\frac{\alpha}{R_5} \partial_5
    & -\partial_5^2 - \partial_6^2 + \frac{\alpha^2}{4 R_5^2}
  \end{array}
  \right)
  \left(
  \begin{array}{c}
    A_{\mpm}^{\mu \, 17} \\ A_{\mmp}^{\mu \, 36}
  \end{array}
  \right)\\
  &=
  \frac{1}{2} \sum_{n=0}^\infty \sum_{m=1}^\infty
  \left(
  \begin{array}{cc}
    A_\mu^{17(n,m)} & A_\mu^{36(n,m)}
  \end{array}
  \right)
  \left(
  \begin{array}{cc}
    \frac{(n+\frac{1}{2})^2}{R_5^2} + \frac{m^2}{R_6^2} + \frac{\alpha^2}{4 R_5^2} & \frac{(n+\frac{1}{2}) \alpha}{R_5^2} \\
    \frac{(n+\frac{1}{2}) \alpha}{R_5^2} & \frac{(n+\frac{1}{2})^2}{R_5^2} + \frac{m^2}{R_6^2} + \frac{\alpha^2}{4 R_5^2}
  \end{array}
  \right)
  \left(
  \begin{array}{c}
    A^{\mu \, 17(n,m)} \\ A^{\mu \, 36(n,m)}
  \end{array}
  \right). 
  \stepcounter{equation}\tag{\theequation}
\end{align*}
Diagonalizing this mass matrix, we obtain the KK mass eigenvalues. 
\begin{align}
  m_{n,m}^2 =
  \frac{(n+\frac{1}{2}\pm\frac{\alpha}{2})^2}{R_5^2}
  +\frac{m^2}{R_6^2}, 
\end{align}
where $n=0, \cdots, \infty, m=1, \cdots, \infty$. 
The KK masses in $a=19-38, 21-40$ sectors can be similarly obtained. 

The mass matrix in $a=22-43$ sector is 
\begin{align*}
  &\frac{1}{2} \int dx_5 \int dx_6 \ 
  \left(
  \begin{array}{cc}
    A_{\mu\ppp}^{22} & A_{\mu\pmm}^{43}
  \end{array}
  \right)
  \left(
  \begin{array}{cc}
    -\partial_5^2 - \partial_6^2 + \frac{\alpha^2}{4 R_5^2}
    & -\frac{\alpha}{R_5} \partial_5 \\
    \frac{\alpha}{R_5} \partial_5
    & -\partial_5^2 - \partial_6^2 + \frac{\alpha^2}{4 R_5^2}
  \end{array}
  \right)
  \left(
  \begin{array}{c}
    A_{\ppp}^{\mu \, 22} \\ A_{\pmm}^{\mu \, 43}
  \end{array}
  \right)\\
  &=
  \frac{1}{2}
  \frac{\alpha^2}{4 R_5^2}
  A_\mu^{22(0,0)}
  A^{\mu \, 22(0,0)} \\
  &+
  \frac{1}{2} \sum_{n=1}^\infty
  \left(
  \begin{array}{cc}
    A_\mu^{22(n,0)} & A_\mu^{43(n,0)}
  \end{array}
  \right)
  \left(
  \begin{array}{cc}
    \frac{n^2}{R_5^2} + \frac{\alpha^2}{4 R_5^2}
    & -\frac{n \alpha}{R_5^2} \\
    -\frac{n \alpha}{R_5^2}
    & \frac{n^2}{R_5^2} + \frac{\alpha^2}{4 R_5^2}
  \end{array}
  \right)
  \left(
  \begin{array}{c}
    A^{\mu \, 22(n,0)} \\ A^{\mu \, 43(n,0)}
  \end{array}
  \right)\\
  &+
  \frac{1}{2} \sum_{m=1}^\infty
  \left(\frac{m^2}{R_6^2} + \frac{\alpha^2}{4 R_5^2}\right)
  A_\mu^{22(0,m)}
  A^{\mu \, 22(0,m)} \\
  &+
  \frac{1}{2} \sum_{n=1}^\infty \sum_{m=1}^\infty
  \left(
  \begin{array}{cc}
    A_\mu^{22(n,m)} & A_\mu^{43(n,m)}
  \end{array}
  \right)
  \left(
  \begin{array}{cc}
    \frac{n^2}{R_5^2} + \frac{m^2}{R_6^2} + \frac{\alpha^2}{4 R_5^2}
    & -\frac{n \alpha}{R_5^2} \\
    -\frac{n \alpha}{R_5^2}
    & \frac{n^2}{R_5^2} + \frac{m^2}{R_6^2} + \frac{\alpha^2}{4 R_5^2}
  \end{array}
  \right)
  \left(
  \begin{array}{c}
    A^{\mu \, 22(n,m)} \\ A^{\mu \, 43(n,m)}
  \end{array}
  \right). 
  \stepcounter{equation}\tag{\theequation}
\end{align*}
Diagonalizing these mass matrices, we obtain the KK mass eigenvalues. 
\begin{align}
  m_{n,m}^2 =
  \frac{\alpha^2}{4R_5^2}, \
  \frac{(n\pm\frac{\alpha}{2})^2}{R_5^2}, \
  \frac{\alpha^2}{4R_5^2}+\frac{m^2}{R_6^2}, \
  \frac{(n\pm\frac{\alpha}{2})^2}{R_5^2}+\frac{m^2}{R_6^2}, 
\end{align}
where $m, n=1, \cdots, \infty$. 

The mass matrix in $a=23-42$ sector is
\begin{align*}
  &\frac{1}{2} \int dx_5 \int dx_6 \ 
  \left(
  \begin{array}{cc}
    A_{\mu\ppp}^{23} & A_{\mu\pmm}^{42}
  \end{array}
  \right)
  \left(
  \begin{array}{cc}
    -\partial_5^2 - \partial_6^2 + \frac{\alpha^2}{4 R_5^2}
    & \frac{\alpha}{R_5} \partial_5 \\
    -\frac{\alpha}{R_5} \partial_5
    & -\partial_5^2 - \partial_6^2 + \frac{\alpha^2}{4 R_5^2}
  \end{array}
  \right)
  \left(
  \begin{array}{c}
    A_{\ppp}^{\mu \, 23} \\ A_{\pmm}^{\mu \, 42}
  \end{array}
  \right)\\
  &=
  \frac{1}{2}
  \frac{\alpha^2}{4 R_5^2}
  A_\mu^{23(0,0)}
  A^{\mu \, 23(0,0)} \\
  &+
  \frac{1}{2} \sum_{n=1}^\infty
  \left(
  \begin{array}{cc}
    A_\mu^{23(n,0)} & A_\mu^{42(n,0)}
  \end{array}
  \right)
  \left(
  \begin{array}{cc}
    \frac{n^2}{R_5^2} + \frac{\alpha^2}{4 R_5^2}
    & \frac{n \alpha}{R_5^2} \\
    \frac{n \alpha}{R_5^2}
    & \frac{n^2}{R_5^2} + \frac{\alpha^2}{4 R_5^2}
  \end{array}
  \right)
  \left(
  \begin{array}{c}
    A^{\mu \, 23(n,0)} \\ A^{\mu \, 42(n,0)}
  \end{array}
  \right)\\
  &+
  \frac{1}{2} \sum_{m=1}^\infty
  \left(\frac{m^2}{R_6^2} + \frac{\alpha^2}{4 R_5^2}\right)
  A_\mu^{23(0,m)}
  A^{\mu \, 23(0,m)} \\
  &+
  \frac{1}{2} \sum_{n=1}^\infty \sum_{m=1}^\infty
  \left(
  \begin{array}{cc}
    A_\mu^{23(n,m)} & A_\mu^{42(n,m)}
  \end{array}
  \right)
  \left(
  \begin{array}{cc}
    \frac{n^2}{R_5^2} + \frac{m^2}{R_6^2} + \frac{\alpha^2}{4 R_5^2}
    & \frac{n \alpha}{R_5^2} \\
    \frac{n \alpha}{R_5^2}
    & \frac{n^2}{R_5^2} + \frac{m^2}{R_6^2} + \frac{\alpha^2}{4 R_5^2}
  \end{array}
  \right)
  \left(
  \begin{array}{c}
    A^{\mu \, 23(n,m)} \\ A^{\mu \, 42(n,m)}
  \end{array}
  \right)
  \stepcounter{equation}\tag{\theequation}
\end{align*}
Diagonalizing this matrix, we obtain the KK mass eigenvalues.  
\begin{align}
  m_{n,m}^2 =
  \frac{\alpha^2}{4R_5^2}, \
  \frac{(n\pm\frac{\alpha}{2})^2}{R_5^2}, \
  \frac{\alpha^2}{4R_5^2}+\frac{m^2}{R_6^2}, \
  \frac{(n\pm\frac{\alpha}{2})^2}{R_5^2}+\frac{m^2}{R_6^2},
\end{align}
where $m, n=1, \cdots, \infty$. 

The mass matrix in $a=33-47$ sector is 
\begin{align*}
  &\frac{1}{2} \int dx_5 \int dx_6 \ 
  \left(
  \begin{array}{cc}
    A_{\mu\pmp}^{33} & A_{\mu\ppm}^{47}
  \end{array}
  \right)
  \left(
  \begin{array}{cc}
    -\partial_5^2 - \partial_6^2 + \frac{\alpha^2}{4 R_5^2}
    & -\frac{\alpha}{R_5} \partial_5 \\
    \frac{\alpha}{R_5} \partial_5
    & -\partial_5^2 - \partial_6^2 + \frac{\alpha^2}{4 R_5^2}
  \end{array}
  \right)
  \left(
  \begin{array}{c}
    A_{\pmp}^{\mu \, 33} \\ A_{\ppm}^{\mu \, 47}
  \end{array}
  \right)\\
  &=
  \frac{1}{2} \sum_{n=0}^\infty
  \left(
  \begin{array}{cc}
    A_\mu^{33(n,0)} & A_\mu^{47(n,0)}
  \end{array}
  \right)
  \left(
  \begin{array}{cc}
    \frac{(n+\frac{1}{2})^2}{R_5^2} + \frac{\alpha^2}{4 R_5^2}
    & \frac{(n+\frac{1}{2}) \alpha}{R_5^2} \\
    \frac{(n+\frac{1}{2}) \alpha}{R_5^2}
    & \frac{(n+\frac{1}{2})^2}{R_5^2} + \frac{\alpha^2}{4 R_5^2}
  \end{array}
  \right)
  \left(
  \begin{array}{c}
    A^{\mu \, 33(n,0)} \\ A^{\mu \, 47(n,0)}
  \end{array}
  \right) \\
  &+
  \frac{1}{2} \sum_{n=0}^\infty \sum_{m=1}^\infty
  \left(
  \begin{array}{cc}
    A_\mu^{33(n,m)} & A_\mu^{47(n,m)}
  \end{array}
  \right)
  \left(
  \begin{array}{cc}
    \frac{(n+\frac{1}{2})^2}{R_5^2} + \frac{m^2}{R_6^2} + \frac{\alpha^2}{4 R_5^2} & \frac{(n+\frac{1}{2}) \alpha}{R_5^2} \\
    \frac{(n+\frac{1}{2}) \alpha}{R_5^2} & \frac{(n+\frac{1}{2})^2}{R_5^2} + \frac{m^2}{R_6^2} + \frac{\alpha^2}{4 R_5^2}
  \end{array}
  \right)
  \left(
  \begin{array}{c}
    A^{\mu \, 33(n,m)} \\ A^{\mu \, 47(n,m)}
  \end{array}
  \right)
  \stepcounter{equation}\tag{\theequation}
\end{align*}
Diagonalizing this mass matrix, we obtain the KK mass eigenvalues. 
\begin{align}
  m_{n,m}^2 =
  \frac{(n+\frac{1}{2}\pm\frac{\alpha}{2})^2}{R_5^2}, \
  \frac{(n+\frac{1}{2}\pm\frac{\alpha}{2})^2}{R_5^2}
  +\frac{m^2}{R_6^2},
\end{align}
where $n=0, \cdots, \infty, m=1, \cdots, \infty$. 
The KK masses in $a=34-46$ sector can be similarly obtained. 

The KK masses of the 4D gauge boson dependent on the VEV $\alpha$ are summarized below. 
\begin{align}
  m_{n,m}^2=&
  \frac{\alpha^2}{R_5^2}, \quad 
  \frac{\alpha^2}{4R_5^2} \times 2, \quad 
  \frac{(n\pm\alpha)^2}{R_5^2}, \quad
  \frac{(n\pm\frac{\alpha}{2})^2}{R_5^2} \times 2, \quad
  \frac{(\tilde{n}+\frac{1}{2}\pm\frac{\alpha}{2})^2}{R_5^2} \times 2, \notag \\    
  &
  \frac{\alpha^2}{R_5^2}+\frac{m^2}{R_6^2}, \quad
  \left(\frac{(\frac{\alpha}{2})^2}{R_5^2}+\frac{m^2}{R_6^2}\right) \times 2,\quad  
  \frac{(n\pm\alpha)^2}{R_5^2}+\frac{m^2}{R_6^2}, \notag \\
 & \left(\frac{(n\pm\frac{\alpha}{2})^2}{R_5^2}+\frac{m^2}{R_6^2}\right) \times 2, \quad
  \left(\frac{(\tilde{n}+\frac{1}{2}\pm\frac{\alpha}{2})^2}{R_5^2}+\frac{m^2}{R_6^2}\right) \times 8, 
\end{align}
where $n=1, \cdots, \infty, \tilde{n}=0, \cdots, \infty, m=1, \cdots, \infty$. 

\subsection{Diagonalization of the mass matrices for $A_5, A_6$}
\label{subsec:56KKmass}
The relevant mass terms for $A_5, A_6$ are the following. 
\begin{align}
  &\frac{1}{2}A_6^a(D_5 D^5 A^6)^a + (\partial_6 A_5)^a(D^5 A^6)^a
  - \frac{1}{2}(\partial_6 A_5)^a(\partial^6 A^5)^a \notag \\
  & \qquad 
  = - \frac{1}{2} A_6^a \left(I^{ab}\partial_5 + f^{a\,44\,b} \frac{\alpha}{R_5}\right)
  \left(I^{bc} \partial_5 + f^{b\,44\,c} \frac{\alpha}{R_5}\right) A^{6 \, c} \notag \\
  & \quad \qquad
  + \frac{1}{2} A_5^a \left(I^{ab} \partial_5 \partial_6 + f^{a\,44\,b}\frac{\alpha}{R_5} \partial_6\right) A^{6\,b}
  + \frac{1}{2}A_6^a \left(I^{ab}\partial_5 \partial_6 + f^{a\,44\,b}\frac{\alpha}{R_5} \partial_6\right) A^{5\,b} \notag \\
  & \quad \qquad
  -\frac{1}{2}A_5^a (I^{ac}\partial_6 \partial_6) A^{5\,c}
  +\cdots. 
\end{align}
The mass matrices in $a=15-24-35-45$ sector are 
{\small
\begin{align*}
  & \frac{1}{2} \int dx_5 \int dx_6 \
  \left(
  \begin{array}{cccc|cccc}
    A_{5\pmm}^{15} & A_{5\pmm}^{24} &
    A_{5\pmm}^{35} & A_{5\ppp}^{45} &
    A_{6\mpp}^{15} & A_{6\mpp}^{24} &
    A_{6\mpp}^{35} & A_{6\mmm}^{45}
  \end{array}
  \right) \\
  & \quad \times
  \left(
  \begin{array}{cccc|cccc}
    -\partial_6^2 & 0 & 0 & 0 &
    \partial_5 \partial_6 & 0 & 0 & -\frac{\alpha}{2 R_5} \partial_6 \\
    0 & -\partial_6^2 & 0 & 0 &
    0 & \partial_5 \partial_6 & 0 & \frac{\alpha}{2\sqrt{3} R_5} \partial_6 \\
    0 & 0 & -\partial_6^2 & 0 &
    0 & 0 & \partial_5 \partial_6 & \sqrt{\frac{2}{3}}\frac{\alpha}{R_5} \partial_6 \\
    0 & 0 & 0 & -\partial_6^2 &
    \frac{m \alpha}{2 R_5 R_6} & -\frac{m \alpha}{2\sqrt{3}R_5 R_6} & -\sqrt{\frac{2}{3}}\frac{m \alpha}{R_5 R_6} & \partial_5 \partial_6 \\
    \hline
    \partial_5 \partial_6 & 0 & 0 & \frac{m \alpha}{2 R_5 R_6} &
    -\partial_5^2 + \frac{\alpha^2}{4 R_5^2} & -\frac{\alpha^2}{4\sqrt{3} R_5^2} 
    & -\frac{\alpha^2}{\sqrt{6} R_5^2} & \frac{\alpha}{R_5} \partial_5 \\
    0 & \partial_5 \partial_6 & 0 & -\frac{m \alpha}{2\sqrt{3}R_5 R_6} &
    -\frac{\alpha^2}{4\sqrt{3} R_5^2} & -\partial_5^2 + \frac{\alpha^2}{12 R_5^2} 
    & \frac{\alpha^2}{3\sqrt{2}R_5^2} & -\frac{\alpha}{\sqrt{3} R_5} \partial_5 \\
    0 & 0 & \partial_5 \partial_6 & -\sqrt{\frac{2}{3}}\frac{m \alpha}{R_5 R_6} &
    -\frac{\alpha^2}{\sqrt{6} R_5^2} & \frac{\alpha^2}{3\sqrt{2}R_5^2} 
    & -\partial_5^2 + \frac{2 \alpha^2}{3 R_5^2} & -2\sqrt{\frac{2}{3}}\frac{\alpha}{R_5} \partial_5 \\
    \frac{\alpha}{2 R_5} \partial_6 & -\frac{\alpha}{2\sqrt{3} R_5} \partial_6 
    & -\sqrt{\frac{2}{3}}\frac{\alpha}{R_5} \partial_6 & \partial_5 \partial_6 &
    -\frac{\alpha}{R_5} \partial_5 & \frac{\alpha}{\sqrt{3} R_5} \partial_5 
    & 2\sqrt{\frac{2}{3}}\frac{\alpha}{R_5} \partial_5 & -\partial_5^2 + \frac{\alpha^2}{R_5^2}
  \end{array}
  \right) \\
  & \quad \times
  \left(
  \begin{array}{c}
    A_{\pmm}^{5\,15}\\
    A_{\pmm}^{5\,24}\\
    A_{\pmm}^{5\,35}\\ 
    A_{\pmm}^{5\,45}\\
    \hline
    A_{\mpp}^{6\,15}\\ 
    A_{\mpp}^{6\,24}\\
    A_{\mpp}^{6\,35}\\ 
    A_{\mmm}^{6\,45}
  \end{array}
  \right) \\
  &=
  \frac{1}{2} \sum_{m=1}^\infty
  \left(
  \begin{array}{c|ccc}
    A_5^{45(0,m)} & A_6^{15(0,m)} &
    A_6^{24(0,m)} & A_6^{35(0,m)}
  \end{array}
  \right) \\
  & \quad \times
  \left(
  \begin{array}{c|ccc}
    \frac{m^2}{R_6^2} & \frac{m \alpha}{2 R_5 R_6} & -\frac{m \alpha}{2\sqrt{3}R_5 R_6} & -\sqrt{\frac{2}{3}}\frac{m \alpha}{R_5 R_6} \\
    \hline
    \frac{m \alpha}{2 R_5 R_6} & \frac{\alpha^2}{4 R_5^2} & -\frac{\alpha^2}{4\sqrt{3} R_5^2} & -\frac{\alpha^2}{\sqrt{6} R_5^2} \\
    -\frac{m \alpha}{2\sqrt{3}R_5 R_6} & -\frac{\alpha^2}{4\sqrt{3} R_5^2} & \frac{\alpha^2}{12 R_5^2} & \frac{\alpha^2}{3\sqrt{2}R_5^2} \\
    -\sqrt{\frac{2}{3}}\frac{m \alpha}{R_5 R_6} & -\frac{\alpha^2}{\sqrt{6} R_5^2} & \frac{\alpha^2}{3\sqrt{2}R_5^2} & \frac{2 \alpha^2}{3 R_5^2}
  \end{array}
  \right)
  \left(
  \begin{array}{c} 
    A^{5\,45(0,m)}\\
    \hline
    A^{6\,15(0,m)}\\ 
    A^{6\,24(0,m)}\\
    A^{6\,35(0,m)}
  \end{array}
  \right)\\
  &+
  \frac{1}{2} \sum_{n=1}^\infty \sum_{m=1}^\infty
  \left(
  \begin{array}{cccc|cccc}
    A_5^{15(n,m)} & A_5^{24(n,m)} &
    A_5^{35(n,m)} & A_5^{45(n,m)} &
    A_6^{15(n,m)} & A_6^{24(n,m)} &
    A_6^{35(n,m)} & A_6^{45(n,m)}
  \end{array}
  \right) \\
  & \quad \times
  \left(
  \begin{array}{cccc|cccc}
    \frac{m^2}{R_6^2} & 0 & 0 & 0 &
    -\frac{nm}{R_5 R_6} & 0 & 0 & -\frac{m \alpha}{2 R_5 R_6} \\
    0 & \frac{m^2}{R_6^2} & 0 & 0 &
    0 & -\frac{nm}{R_5 R_6} & 0 & \frac{m \alpha}{2\sqrt{3} R_5 R_6} \\
    0 & 0 & \frac{m^2}{R_6^2} & 0 &
    0 & 0 & -\frac{nm}{R_5 R_6} & \sqrt{\frac{2}{3}}\frac{m \alpha}{R_5 R_6} \\
    0 & 0 & 0 & \frac{m^2}{R_6^2} &
    \frac{m \alpha}{2 R_5 R_6} & -\frac{m \alpha}{2\sqrt{3}R_5 R_6} & -\sqrt{\frac{2}{3}}\frac{m \alpha}{R_5 R_6} & \frac{nm}{R_5 R_6} \\
    \hline
    - \frac{nm}{R_5 R_6} & 0 & 0 & \frac{m \alpha}{2 R_5 R_6} &
    \frac{n^2}{R_5^2} + \frac{\alpha^2}{4 R_5^2} & -\frac{\alpha^2}{4\sqrt{3} R_5^2} & -\frac{\alpha^2}{\sqrt{6} R_5^2} & \frac{n \alpha}{R_5^2} \\
    0 & - \frac{nm}{R_5 R_6} & 0 & -\frac{m \alpha}{2\sqrt{3}R_5 R_6} &
    -\frac{\alpha^2}{4\sqrt{3} R_5^2} & \frac{n^2}{R_5^2} + \frac{\alpha^2}{12 R_5^2} & \frac{\alpha^2}{3\sqrt{2}R_5^2} & -\frac{n \alpha}{\sqrt{3} R_5^2} \\
    0 & 0 & - \frac{nm}{R_5 R_6} & -\sqrt{\frac{2}{3}}\frac{m \alpha}{R_5 R_6} &
    -\frac{\alpha^2}{\sqrt{6} R_5^2} & \frac{\alpha^2}{3\sqrt{2}R_5^2} & \frac{n^2}{R_5^2} + \frac{2 \alpha^2}{3 R_5^2} & -2\sqrt{\frac{2}{3}}\frac{n \alpha}{R_5^2} \\
    -\frac{m \alpha}{2 R_5 R_6} & \frac{m \alpha}{2\sqrt{3} R_5 R_6} & \sqrt{\frac{2}{3}}\frac{m \alpha}{R_5 R_6} & \frac{nm}{R_5 R_6} &
    \frac{n \alpha}{R_5^2} & -\frac{n \alpha}{\sqrt{3} R_5^2} & -2\sqrt{\frac{2}{3}}\frac{n \alpha}{R_5^2} & \frac{n^2}{R_5^2} + \frac{\alpha^2}{R_5^2}
  \end{array}
  \right) 
\end{align*}
\begin{align*}
  \quad \times
  \left(
  \begin{array}{c}
    A^{5\,15(n,m)}\\
    A^{5\,24(n,m)}\\
    A^{5\,35(n,m)}\\ 
    A^{5\,45(n,m)}\\
    \hline
    A^{6\,15(n,m)}\\ 
    A^{6\,24(n,m)}\\
    A^{6\,35(n,m)}\\ 
    A^{6\,45(n,m)}
  \end{array}
  \right)
  \stepcounter{equation}\tag{\theequation}
\end{align*}
}
Diagonalizing these mass matrices, we obtain the KK mass eigenvalues.  
\begin{align}
  m_{n,m}^2 =
  0,\ 0,\ 0,\ 0,\ 0,\ 0,\ 0,\
  \frac{\alpha^2}{R_5^2}+\frac{m^2}{R_6^2},\
  \frac{n^2}{R_5^2}+\frac{m^2}{R_6^2},\
  \frac{n^2}{R_5^2}+\frac{m^2}{R_6^2},\
  \frac{(n\pm\alpha)^2}{R_5^2}+\frac{m^2}{R_6^2},
\end{align}
where $m, n=1, \cdots, \infty$. 

The mass matrices in $a=16-37$ sector are
\begin{align*}
  &\frac{1}{2} \int dx_5 \int dx_6 \
  \left(
  \begin{array}{cc|cc}
    A_{5\mmp}^{16} & A_{5\mpm}^{37} &
    A_{6\ppm}^{16} & A_{6\pmp}^{37}
  \end{array}
  \right) \\
  & \quad \times
  \left(
  \begin{array}{cc|cc}
    -\partial_6^2 & 0 & \partial_5 \partial_6 & \frac{\alpha}{2 R_5} \partial_6 \\
    0 & -\partial_6^2 & - \frac{\alpha}{2 R_5} \partial_6 & \partial_5 \partial_6 \\
    \hline
    \partial_5 \partial_6 & \frac{\alpha}{2 R_5} \partial_6 & -\partial_5^2 + \frac{\alpha^2}{4 R_5^2} & -\frac{\alpha}{R_5} \partial_5 \\
    -\frac{\alpha}{2 R_5} \partial_6 & \partial_5 \partial_6 & \frac{\alpha}{R_5} \partial_5 & -\partial_5^2 + \frac{\alpha^2}{4 R_5^2}
  \end{array}
  \right)
  \left(
  \begin{array}{c}
    A_{\mmp}^{5\,16} \\ A_{\mpm}^{5\,37} \\
    \hline
    A_{\ppm}^{6\,16} \\ A_{\pmp}^{6\,37}
  \end{array}
  \right) \\
  &=
  \frac{1}{2} \sum_{n=0}^\infty
  \left(
  \begin{array}{cc}
    A_6^{16(n,0)} & A_6^{37(n,0)}
  \end{array}
  \right)
  \left(
  \begin{array}{cc}
    \frac{(n+\frac{1}{2})^2}{R_5^2} + \frac{\alpha^2}{4 R_5^2} &-\frac{(n+\frac{1}{2}) \alpha}{R_5^2} \\
    - \frac{(n+\frac{1}{2}) \alpha}{R_5^2} & \frac{(n+\frac{1}{2})^2}{R_5^2} + \frac{\alpha^2}{4 R_5^2}
  \end{array}
  \right)
  \left(
  \begin{array}{c}
    A^{6\,16(n,0)} \\ A^{6\,37(n,0)}
  \end{array}
  \right) \\
  &+
  \frac{1}{2} \sum_{n=0}^\infty \sum_{m=1}^\infty
  \left(
  \begin{array}{cc|cc}
    A_5^{16(n,m)} & A_5^{37(n,m)} &
    A_6^{16(n,m)} & A_6^{37(n,m)}
  \end{array}
  \right) \\
  & \quad \times
  \left(
  \begin{array}{cc|cc}
    \frac{m^2}{R_6^2} & 0 & \frac{(n+\frac{1}{2})m}{R_5 R_6} & -\frac{m \alpha}{2 R_5 R_6} \\
    0 & \frac{m^2}{R_6^2} & \frac{m \alpha}{2 R_5 R_6} & -\frac{(n+\frac{1}{2})m}{R_5 R_6} \\
    \hline
    \frac{(n+\frac{1}{2})m}{R_5 R_6} & \frac{m \alpha}{2 R_5 R_6} 
    & \frac{(n+\frac{1}{2})^2}{R_5^2} + \frac{\alpha^2}{4 R_5^2} & -\frac{(n+\frac{1}{2}) \alpha}{R_5^2} \\
    -\frac{m \alpha}{2 R_5 R_6} & -\frac{(n+\frac{1}{2})m}{R_5 R_6} & - \frac{(n+\frac{1}{2}) \alpha}{R_5^2} 
    & \frac{(n+\frac{1}{2})^2}{R_5^2} + \frac{\alpha^2}{4 R_5^2}
  \end{array}
  \right)
  \left(
  \begin{array}{c}
    A^{5\,16(n,m)} \\ A^{5\,37(n,m)} \\
    \hline
    A^{6\,16(n,m)} \\ A^{6\,37(n,m)}
  \end{array}
  \right). 
  \stepcounter{equation}\tag{\theequation}
\end{align*}
Diagonalizing these mass matrices, we obtain the Kk mass eigenvalues. 
\begin{align}
  m_{n,m}^2 =
  0,\ 0,\
  \frac{(n+\frac{1}{2}\pm\frac{\alpha}{2})^2}{R_5^2},\ 
  \frac{(n+\frac{1}{2}\pm\frac{\alpha}{2})^2}{R_5^2}
  +\frac{m^2}{R_6^2},
\end{align}
where $n=0, \cdots, \infty, m=1, \cdots, \infty$. 
The KK masses in $a=18-39, 20-41$ sectors can be similarly obtained. 

The mass matrices in $a=17-36$ sector are
\begin{align*}
  &\frac{1}{2} \int dx_5 \int dx_6 \
  \left(
  \begin{array}{cc|cc}
    A_{5\mmp}^{17} & A_{5\mpm}^{36} & A_{6\ppm}^{17} & A_{6\pmp}^{36}
  \end{array}
  \right) \\
  & \quad \times
  \left(
  \begin{array}{cc|cc}
    -\partial_6^2 & 0 & \partial_5 \partial_6 & - \frac{\alpha}{2 R_5} \partial_6 \\
    0 & -\partial_6^2 & \frac{\alpha}{2 R_5} \partial_6 & \partial_5 \partial_6 \\
    \hline
    \partial_5 \partial_6 & -\frac{\alpha}{2 R_5} \partial_6 & -\partial_5^2 + \frac{\alpha^2}{4 R_5^2} & -\frac{\alpha}{R_5} \partial_5 \\
    \frac{\alpha}{2 R_5} \partial_6 & \partial_5 \partial_6 & \frac{\alpha}{R_5} \partial_5 & -\partial_5^2 + \frac{\alpha^2}{4 R_5^2}
  \end{array}
  \right)
  \left(
  \begin{array}{c}
    A_{\mmp}^{5\,17} \\
    A_{\mpm}^{5\,36} \\
    \hline
    A_{\ppm}^{6\,17} \\
    A_{\pmp}^{6\,36}
  \end{array}
  \right)\\
  &=
  \frac{1}{2} \sum_{n=0}^\infty
  \left(
  \begin{array}{cc}
    A_6^{17(n,0)} & A_6^{36(n,0)}
  \end{array}
  \right)
  \left(
  \begin{array}{cc}
    \frac{(n+\frac{1}{2})^2}{R_5^2} + \frac{\alpha^2}{4 R_5^2} &\frac{(n+\frac{1}{2}) \alpha}{R_5^2} \\
    \frac{(n+\frac{1}{2}) \alpha}{R_5^2} & \frac{(n+\frac{1}{2})^2}{R_5^2} + \frac{\alpha^2}{4 R_5^2}
  \end{array}
  \right)
  \left(
  \begin{array}{c}
    A^{6\,17(n,0)} \\ A^{6\,36(n,0)}
  \end{array}
  \right) \\
  &+
  \frac{1}{2} \sum_{n=0}^\infty \sum_{m=1}^\infty
  \left(
  \begin{array}{cc|cc}
    A_5^{17(n,m)} & A_5^{36(n,m)} & A_6^{17(n,m)} & A_6^{36(n,m)}
  \end{array}
  \right) \\
  & \quad \times
  \left(
  \begin{array}{cc|cc}
    \frac{m^2}{R_6^2} & 0 & \frac{(n+\frac{1}{2})m}{R_5 R_6} & \frac{m \alpha}{2 R_5 R_6} \\
    0 & \frac{m^2}{R_6^2} & -\frac{m \alpha}{2 R_5 R_6} & -\frac{(n+\frac{1}{2})m}{R_5 R_6} \\
    \hline
    \frac{(n+\frac{1}{2})m}{R_5 R_6} & -\frac{m \alpha}{2 R_5 R_6} 
    & \frac{(n+\frac{1}{2})^2}{R_5^2} + \frac{\alpha^2}{4 R_5^2} & \frac{(n+\frac{1}{2})\alpha}{R_5^2} \\
    \frac{m \alpha}{2 R_5 R_6} & -\frac{(n+\frac{1}{2})m}{R_5 R_6} 
    & \frac{(n+\frac{1}{2}) \alpha}{R_5^2} & \frac{(n+\frac{1}{2})^2}{R_5^2} + \frac{\alpha^2}{4 R_5^2}
  \end{array}
  \right)
  \left(
  \begin{array}{c}
    A^{5\,17(n,m)} \\
    A^{5\,36(n,m)} \\
    \hline
    A^{6\,17(n,m)} \\
    A^{6\,36(n,m)}
  \end{array}
  \right). 
  \stepcounter{equation}\tag{\theequation}
\end{align*}
Diagonalizing these mass matrices, we obtain the KK mass eigenvalues. 
\begin{align}
  m_{n,m}^2 =
  0,\ 0,\
  \frac{(n+\frac{1}{2}\pm\frac{\alpha}{2})^2}{R_5^2}, \
  \frac{(n+\frac{1}{2}\pm\frac{\alpha}{2})^2}{R_5^2}
  +\frac{m^2}{R_6^2}, 
\end{align}
where $n=0, \cdots, \infty, m=1, \cdots, \infty$. 
The KK masses in $a=19-38, 21-40$ sectors can be similarly obtained. 

The mass matrices in $a=22-43$ sectors are
\begin{align*}
  &\frac{1}{2} \int dx_5 \int dx_6 \
  \left(
  \begin{array}{cc|cc}
    A_{5\pmm}^{22} & A_{5\ppp}^{43} & A_{6\mpp}^{22} & A_{6\mmm}^{43}
  \end{array}
  \right) \\
  & \quad \times
  \left(
  \begin{array}{cc|cc}
    - \partial_6^2 & 0 & \partial_5 \partial_6 & \frac{\alpha}{2 R_5} \partial_6 \\
    0 & - \partial_6^2 & -\frac{\alpha}{2 R_5} \partial_6 & \partial_5 \partial_6 \\
    \hline
    \partial_5 \partial_6 & \frac{\alpha}{2 R_5} \partial_6 & -\partial_5^2 + \frac{\alpha^2}{4 R_5^2} & -\frac{\alpha}{R_5} \partial_5 \\
    -\frac{\alpha}{2 R_5} \partial_6 & \partial_5 \partial_6 & \frac{\alpha}{R_5} \partial_5 & \frac{\alpha^2}{4 R_5^2}
  \end{array}
  \right)
  \left(
  \begin{array}{c}
    A_{\pmm}^{5\,22} \\
    A_{\ppp}^{5\,43} \\
    \hline
    A_{\mpp}^{6\,22} \\
    A_{\mmm}^{6\,43}
  \end{array}
  \right) \\
  &=
  \frac{1}{2} \sum_{m=1}^\infty
  \left(
  \begin{array}{c|c}
    A_5^{43(0,m)} & A_6^{22(0,m)}
  \end{array}
  \right)
  \left(
  \begin{array}{c|c}
    \frac{m^2}{R_6^2} & -\frac{m \alpha}{2 R_5 R_6} \\
    \hline
    -\frac{m \alpha}{2 R_5 R_6} & \frac{n^2}{R_5^2} + \frac{\alpha^2}{4 R_5^2}
  \end{array}
  \right)
  \left(
  \begin{array}{c}
    A^{5\,43(0,m)} \\
    \hline
    A^{6\,22(0,m)}
  \end{array}
  \right)\\
  &+
  \frac{1}{2} \sum_{n=1}^\infty \sum_{m=1}^\infty
  \left(
  \begin{array}{cc|cc}
    A_5^{22(n,m)} & A_5^{43(n,m)} & A_6^{22(n,m)} & A_6^{43(n,m)}
  \end{array}
  \right) \\
  & \quad \times
  \left(
  \begin{array}{cc|cc}
    \frac{m^2}{R_6^2} & 0 & -\frac{nm}{R_5 R_6} & \frac{m \alpha}{2 R_5 R_6} \\
    0 & \frac{m^2}{R_6^2} & -\frac{m \alpha}{2 R_5 R_6} & \frac{nm}{R_5 R_6} \\
    \hline
    -\frac{nm}{R_5 R_6} & -\frac{m \alpha}{2 R_5 R_6} & \frac{n^2}{R_5^2} + \frac{\alpha^2}{4 R_5^2} & -\frac{n \alpha}{R_5^2} \\
    \frac{m \alpha}{2 R_5 R_6} & \frac{nm}{R_5 R_6} & -\frac{n \alpha}{R_5^2} & \frac{n^2}{R_5^2} + \frac{\alpha^2}{4 R_5^2}
  \end{array}
  \right)
  \left(
  \begin{array}{c}
    A^{5\,22(n,m)} \\
    A^{5\,43(n,m)} \\
    \hline
    A^{6\,22(n,m)} \\
    A^{6\,43(n,m)}
  \end{array}
  \right). 
  \stepcounter{equation}\tag{\theequation}
\end{align*}
Diagonalizing these mass matrices, we obtain the KK mass eigenvalues. 
\begin{align}
  m_{n,m}^2 =
  0,\ 0,\ 0,\
  \frac{\alpha^2}{4R_5^2}+\frac{m^2}{R_6^2},\
  \frac{(n\pm\frac{\alpha}{2})^2}{R_5^2}+\frac{m^2}{R_6^2}, 
\end{align}
where $m, n=1, \cdots, \infty$. 
%

The mass matrices in $a=23-42$ sector is
\begin{align*}
  &\frac{1}{2} \int dx_5 \int dx_6 \
  \left(
  \begin{array}{cc|cc}
    A_{5\pmm}^{23} & A_{5\ppp}^{42} & A_{6\mpp}^{23} & A_{6\mmm}^{42}
  \end{array}
  \right) \\
  & \quad \times
  \left(
  \begin{array}{cc|cc}
    -\partial_6^2 & 0 & \partial_5 \partial_6 & -\frac{\alpha}{2 R_5} \partial_6 \\
    0 & -\partial_6^2 & -\frac{\alpha}{2 R_5} \partial_6 & \partial_5 \partial_6 \\
    \hline
    -\partial_5 \partial_6 & -\frac{\alpha}{2 R_5} \partial_6 & -\partial_5^2 + \frac{\alpha^2}{4 R_5^2} & -\frac{\alpha}{R_5} \partial_5 \\
    \frac{\alpha}{2 R_5} \partial_6 & \partial_5 \partial_6 & \frac{\alpha}{R_5} \partial_5 & -\partial_5^2 + \frac{\alpha^2}{4 R_5^2}
  \end{array}
  \right)
  \left(
  \begin{array}{c}
    A_{\pmm}^{5\,23} \\
    A_{\ppp}^{5\,42} \\
    \hline
    A_{\mpp}^{6\,23} \\
    A_{\mmm}^{6\,42}
  \end{array}
  \right) \\
  &= \frac{1}{2} \sum_{m=1}^\infty
  \left(
  \begin{array}{c|c}
    A_5^{42(0,m)} & A_6^{23(0,m)}
  \end{array}
  \right)
  \left(
  \begin{array}{c|c}
    \frac{m^2}{R_6^2} & \frac{m \alpha}{2 R_5 R_6} \\
    \hline
    \frac{m \alpha}{2 R_5 R_6} & \frac{\alpha^2}{4 R_5^2}
  \end{array}
  \right)
  \left(
  \begin{array}{c}
    A^{5\,42(0,m)} \\
    \hline
    A^{6\,23(0,m)}
  \end{array}
  \right)\\
  &+
  \frac{1}{2} \sum_{n=1}^\infty \sum_{m=1}^\infty
  \left(
  \begin{array}{cc|cc}
    A_5^{23(n,m)} & A_5^{42(n,m)} & A_6^{23(n,m)} & A_6^{42(n,m)}
  \end{array}
  \right) \\
  & \quad \times
  \left(
  \begin{array}{cc|cc}
    \frac{m^2}{R_6^2} & 0 & -\frac{nm}{R_5 R_6} & -\frac{m \alpha}{2 R_5 R_6} \\
    0 & \frac{m^2}{R_6^2} & \frac{m \alpha}{2 R_5 R_6} & \frac{nm}{R_5 R_6} \\
    \hline
    -\frac{nm}{R_5 R_6} & \frac{m \alpha}{2 R_5 R_6} & \frac{n^2}{R_5^2} + \frac{\alpha^2}{4 R_5^2} & \frac{n \alpha}{R_5^2} \\
    -\frac{m \alpha}{2 R_5 R_6} & \frac{nm}{R_5 R_6} & \frac{n \alpha}{R_5^2} & \frac{n^2}{R_5^2} + \frac{\alpha^2}{4 R_5^2}
  \end{array}
  \right)
  \left(
  \begin{array}{c}
    A^{5\,23(n,m)} \\
    A^{5\,42(n,m)} \\
    \hline
    A^{6\,23(n,m)} \\
    A^{6\,42(n,m)}
  \end{array}
  \right).
  \stepcounter{equation}\tag{\theequation}
\end{align*}
Diagonalizing these mass matrices, we obtain the Kk mass eigenvalues.  
\begin{align}
  m_{n,m}^2 =
  0,\ 0,\ 0,\
  \frac{\alpha^2}{4R_5^2}+\frac{m^2}{R_6^2},\
  \frac{(n\pm\frac{\alpha}{2})^2}{R_5^2}+\frac{m^2}{R_6^2},
\end{align}
where $m, n=1, \cdots, \infty$. 
%

The mass matrices in $a=33-47$ sector are
\begin{align}
  &\frac{1}{2} \int dx_5 \int dx_6 \
  \left(
  \begin{array}{cc|cc}
    A_{5\ppm}^{33} & A_{5\pmp}^{47} & A_{6\mmp}^{33} & A_{6\mpm}^{47}
  \end{array}
  \right) \notag \\
  & \quad \times
  \left(
  \begin{array}{cc|cc}
    -\partial_6^2 & 0 & \partial_5 \partial_6 & \frac{\alpha}{2 R_5} \partial_6 \\
    0 & -\partial_6^2 & -\frac{\alpha}{2 R_5} \partial_6 & \partial_5 \partial_6 \\
    \hline
    \partial_5 \partial_6 & \frac{\alpha}{2 R_5} \partial_6 & -\partial_5^2 + \frac{\alpha^2}{4 R_5^2} & -\frac{\alpha}{R_5} \partial_5 \\
    -\frac{\alpha}{2 R_5} \partial_6 & \partial_5 \partial_6 & \frac{\alpha}{R_5} \partial_5 & -\partial_5^2 + \frac{\alpha^2}{4 R_5^2}
  \end{array}
  \right)
  \left(
  \begin{array}{c}
    A_{\ppm}^{5\,33} \\
    A_{\pmp}^{5\,47} \\
    \hline
    A_{\mmp}^{6\,33} \\
    A_{\mpm}^{6\,47}
  \end{array}
  \right) \notag \\
  &=
  \frac{1}{2} \sum_{n=0}^\infty \sum_{m=1}^\infty
  \left(
  \begin{array}{cc|cc}
    A_5^{33(n,m)} & A_5^{47(n,m)} & A_6^{33(n,m)} & A_6^{47(n,m)}
  \end{array}
  \right) \notag \\
  & \quad \times
  \left(
  \begin{array}{cc|cc}
    \frac{m^2}{R_6^2} & 0 & -\frac{(n+\frac{1}{2})m}{R_5 R_6} & \frac{m \alpha}{2 R_5 R_6} \\
    0 & \frac{m^2}{R_6^2} & -\frac{m \alpha}{2 R_5 R_6} & -\frac{(n+\frac{1}{2})m}{R_5 R_6} \\
    \hline
    -\frac{(n+\frac{1}{2})m}{R_5 R_6} & -\frac{m \alpha}{2 R_5 R_6} 
    & \frac{(n+\frac{1}{2})^2}{R_5^2} + \frac{\alpha^2}{4 R_5^2} & \frac{(n+\frac{1}{2})\alpha}{R_5^2} \\
    \frac{m \alpha}{2 R_5 R_6} & -\frac{(n+\frac{1}{2})m}{R_5 R_6} & \frac{(n+\frac{1}{2})\alpha}{R_5^2} 
    & \frac{(n+\frac{1}{2})^2}{R_5^2} + \frac{\alpha^2}{4 R_5^2}
  \end{array}
  \right)
  \left(
  \begin{array}{c}
    A^{5\,33(n,m)} \\
    A^{5\,47(n,m)} \\
    \hline
    A^{6\,33(n,m)} \\
    A^{6\,47(n,m)}
  \end{array}
  \right). 
  \stepcounter{equation}\tag{\theequation}
\end{align}
Diagonalizing these mass matrices, we obtain the KK mass eigenvalues. 
\begin{align}
  m_{n,m}^2 =
  0,\ 0,\
  \frac{(n+\frac{1}{2}\pm\frac{\alpha}{2})^2}{R_5^2}
  +\frac{m^2}{R_6^2},
 \end{align}
where $n=0, \cdots, \infty, m=1, \cdots, \infty$. 
The KK masses in $a=34-46$ sector can be similarly obtained. 

The KK masses of $A_5, A_6$ dependent on the VEV $\alpha$ are summarized below. 
\begin{align}
  m_{n,m}^2=&
  \left(\frac{(\tilde{n}+\frac{1}{2}\pm\frac{\alpha}{2})^2}{R_5^2}\right) \times 6, \quad 
  \frac{\alpha^2}{R_5^2}+\frac{m^2}{R_6^2}, \quad 
  \left(\frac{\alpha^2}{4R_5^2}+\frac{m^2}{R_6^2}\right) \times 2, \quad 
  \frac{(n\pm\alpha)^2}{R_5^2}+\frac{m^2}{R_6^2}, \notag \\
 &  \left(\frac{(n\pm\frac{\alpha}{2})^2}{R_5^2}+\frac{m^2}{R_6^2}\right) \times 2, \quad
  \left(\frac{(\tilde{n}+\frac{1}{2}\pm\frac{\alpha}{2})^2}{R_5^2}
  +\frac{m^2}{R_6^2}\right) \times 8,
\end{align}
where $n=1, \cdots, \infty, \tilde{n}=0, \cdots, \infty, m=1, \cdots, \infty$. 

The KK mass spectrum of fermions can be obtained similar to those of the 6D gauge boson, 
but we do not repeat the similar calculations since the KK mass spectrum can be easily found 
by group theoretical considerations described in the main text. 

\section{Calculation of one-loop effective potential}
\label{sec:veff}


In this section, the calculation of the one-loop effective potential is discussed in some detail 
by taking the 4D gauge field contributions, for example. 
\begin{align*}
  V_{\rm eff}^\mu
  &= \frac{4}{2(2 \pi)^2 R_5 R_6} \int \frac{d^4 p_E}{(2 \pi)^4} \\
  &\qquad \times
  \left\{
    \log (p_E^2 + \frac{\alpha^2}{R_5^2})
    + 2 \log (p_E^2 + \frac{(\frac{\alpha}{2})^2}{R_5^2})
    \right.\\
    &\quad \qquad + \sum_{n=1}^\infty \left[
    \log (p_E^2 + \frac{(n \pm \alpha)^2}{R_5^2})
    + 2 \log (p_E^2 + \frac{(n \pm \frac{\alpha}{2})^2}{R_5^2})\right]
    + \sum_{n=0}^\infty
    2 \log (p_E^2 + \frac{(n + \frac{1}{2} \pm \frac{\alpha}{2})^2}{R_5^2}) \\
    &\qquad \quad + \sum_{m=1}^\infty \left[
    \log (p_E^2 + \frac{\alpha^2}{R_5^2} + \frac{m^2}{R_6^2})
    + 2 \log (p_E^2 + \frac{(\frac{\alpha}{2})^2}{R_5^2} + \frac{m^2}{R_6^2}) 
    \right] \\
    &\qquad \quad + \sum_{n=1}^\infty \sum_{m=1}^\infty \left[
    \log (p_E^2 + \frac{(n \pm \alpha)^2}{R_5^2}  + \frac{m^2}{R_6^2}) 
    + 2 \log (p_E^2 + \frac{(n \pm \frac{\alpha}{2})^2}{R_5^2}  + \frac{m^2}{R_6^2})\right] \\
    & \left. \qquad \quad
    + \sum_{n=0}^\infty \sum_{m=1}^\infty
    8 \log (p_E^2 + \frac{(n + \frac{1}{2} \pm \frac{\alpha}{2})^2}{R_5^2}  + \frac{m^2}{R_6^2}) 
  \right\} \\
  &= \frac{2}{(2\pi)^2 R_5 R_6} \int \frac{d^4 p_E}{(2\pi)^4} \\
  & \qquad \times
  \left\{ \left[\sum_{n=-\infty}^\infty \sum_{m=1}^\infty 
    \log(p_E^2 + \left(\frac{n-\alpha}{R_5}\right)^2 + \left(\frac{m}{R_6}\right)^2) 
    + \sum_{n=-\infty}^\infty
    \log(p_E^2 + \left(\frac{n-\alpha}{R_5}\right)^2)\right] \right. \\
    & \quad \qquad
    + 2 \left[\sum_{n=-\infty}^\infty \sum_{m=1}^\infty 
    \log(p_E^2 + \left(\frac{n-\frac{\alpha}{2}}{R_5}\right)^2 + \left(\frac{m}{R_6}\right)^2)
    + \sum_{n=-\infty}^\infty
    \log(p_E^2 + \left(\frac{n-\frac{\alpha}{2}}{R_5}\right)^2) \right] \\
    & \quad \qquad
    + 2 \left[\sum_{n=-\infty}^\infty \sum_{m=1}^\infty
    \log(p_E^2 + \left(\frac{n-\frac{1}{2}-\frac{\alpha}{2}}{R_5}\right)^2 + \left(\frac{m}{R_6}\right)^2)
    + \sum_{n=\infty}^\infty
    \log(p_E^2 + \left(\frac{n-\frac{1}{2}-\frac{\alpha}{2}}{R_5}\right)^2)\right] \\
    & \left. \quad \qquad
    + 6 \sum_{n=-\infty}^\infty \sum_{m=1}^\infty
    \log(p_E^2 + \left(\frac{n-\frac{1}{2}-\frac{\alpha}{2}}{R_5}\right)^2 + \left(\frac{m}{R_6}\right)^2)
  \right\}.
  \label{oneloopgauge}
  \stepcounter{equation}\tag{\theequation}
\end{align*}
In the second equality, the mode summation with respect to $n$ is rewritten to the summation from $-\infty$ to $\infty$.
Here, we define the first term, the second term, the third term and the fourth term in (\ref{oneloopgauge}) as
$V_{\rm eff}^{g\ppp},V_{\rm eff}^{g\pmm},
V_{\rm eff}^{g\pmp}$ and $V_{\rm eff}^{g\mpm}$. 

The partial derivative of the first term of the potential $V_{\rm eff}^{g\ppp}$ with respect to $\alpha$ is calculated as
\begin{align*}
  \pdv{V_{\rm eff}^{g \ppp}}{\alpha}
  &= \frac{-1}{2 \pi^2 R_5^2 R_6} \int \frac{d^4 p_E}{(2\pi)^4} \times \notag \\
&  \left\{\sum_{n=-\infty}^\infty \sum_{m=-\infty}^\infty
  \frac{n-\alpha}{R_5}
  \frac{1}{p_E^2 + \left(\frac{n-\alpha}{R_5}\right)^2 + \left(\frac{m}{R_6}\right)^2} 
  + \sum_{n=-\infty}^\infty
  \frac{n-\alpha}{R_5}
  \frac{1}{p_E^2 + \left(\frac{n-\alpha}{R_5}\right)^2}\right\}.
\end{align*}
By $\Gamma$ function formula 
\begin{equation}
  \frac{\Gamma(s)}{D^s}
  = \int_0^\infty dt \ t^{s-1} e^{-Dt}, 
\end{equation}
it can be rewritten as
\begin{align*}
  \pdv{V_{\rm eff}^{g \ppp}}{\alpha}
  &= \frac{-1}{2 \pi^2 R_5^2 R_6} \int_0^\infty dt
  \int \frac{d^4 p_E}{(2\pi)^4} \ e^{-p_E^2 t} 
  \sum_{n=-\infty}^\infty \frac{n-\alpha}{R_5} e^{-\left(\frac{n-\alpha}{R_5}\right)^2 t}
  \left\{\sum_{m=-\infty}^\infty e^{-\left(\frac{m}{R_6}\right)^2 t} + 1 \right\}. 
\end{align*}
By noticing the Poisson resummation formura and its $t$ derivative,  
\begin{equation}
  \sum_{n=-\infty}^\infty
  e^{-\left(\frac{n + t \alpha}{R}\right)^2}
  = \sum_{n=-\infty}^\infty
  R \sqrt{\frac{\pi}{s}}
  e^{-\frac{(\pi R n)^2}{t}}
  e^{-2 \pi i n \alpha t},
\end{equation}
\begin{equation}
  \sum_{n=-\infty}^\infty
  \frac{n + t \alpha}{R}
  e^{-\left(\frac{n + t \alpha}{R}\right)^2}
  = \sum_{n=-\infty}^\infty
  R^2 \sqrt{\frac{\pi}{s^3}}
  (i \pi n)
  e^{-\frac{(\pi R n)^2}{t}}
  e^{-2 \pi i n \alpha t},
\end{equation}
the partial derivative of the potential can be further rewritten in a useful form. 
\begin{align*}
  \quad
  \pdv{V_{\rm eff}^{g \ppp}}{\alpha}
  &= \frac{-1}{2 \pi^2 R_5^2 R_6} \int_0^\infty dt
  \ \frac{1}{(2\pi)^4} \left(\sqrt{\frac{\pi}{t}}\right)^4 \\
  & \qquad \times
  \sum_{n=-\infty}^\infty R_5^2 \ \sqrt{\frac{\pi}{t^3}} \ (i \pi n)
  \ e^{-\frac{(\pi R_5 n)^2}{t}+2 \pi i n \alpha}
  \left\{\sum_{m=-\infty}^\infty R_6 \ \sqrt{\frac{\pi}{t}}
  \ e^{-\frac{(\pi R_6 m)^2}{t}} + 1 \right\} \\
  &= \frac{-\sqrt{\pi}}{2 \pi^2 R_6} \frac{1}{(4 \pi)^2}
  \int_0^\infty dt \ \frac{1}{t^{\frac{7}{2}}} \ (i \pi) \\
  & \qquad \times
  \sum_{n=1}^\infty ne^{-\frac{(\pi R_5 n)^2}{t}}
  \left(e^{2 \pi i n \alpha} - e^{-2 \pi i n \alpha}\right)
  \left\{2 \sum_{m=1}^\infty R_6 \ \sqrt{\frac{\pi}{t}}
  \ e^{-\frac{(\pi R_6 m)^2}{t}} + R_6 \ \sqrt{\frac{\pi}{t}} + 1\right\} \\
  &= \frac{\sqrt{\pi}}{\pi R_6} \frac{1}{(4 \pi)^2}
  \sum_{n=1}^\infty n \sin(2 \pi n \alpha) \\
  & \qquad \times
  \int_0^\infty \frac{du}{u^2}
  \left\{2 \sqrt{\pi} R_6 \sum_{m=1}^\infty 
  u^4 \ e^{-\left((\pi R_5 n)^2 + (\pi R_6 m)^2\right)u}
  + \sqrt{\pi} R_6 \ 
  u^4 \ e^{-(\pi R_5 n)^2 u}
  + u^{\frac{7}{2}} \ e^{-(\pi R_5 n)^2 u}\right\} \\
  &= \frac{\sqrt{\pi}}{\pi R_6} \frac{1}{(4 \pi)^2}
  \sum_{n=1}^\infty n \sin(2 \pi n \alpha) \times \notag \\
 & \hspace*{10mm}\left\{2 \sqrt{\pi} R_6 \sum_{m=1}^\infty 
  \frac{\Gamma(3)}{\left((\pi R_5 n)^2 + (\pi R_6 m)^2\right)^3}
  + \sqrt{\pi} R_6 \ 
  \frac{\Gamma(3)}{(\pi R_5 n)^6}
  + \frac{\Gamma \left(\frac{5}{2}\right)}{(\pi R_5 n)^5}\right\} \\
  &= \frac{1}{(4\pi)^2} \sum_{n=1}^\infty n \sin(2 \pi n \alpha)
  \left\{\sum_{m=1}^\infty 
  \frac{4}{\left((\pi R_5 n)^2 + (\pi R_6 m)^2\right)^3}
  + \frac{2}{(\pi R_5 n)^6}
  + \frac{3}{4(\pi R_5 n)^5 R_6}\right\}
\end{align*}
In the third equality, the changing of variable $t=1/u$ is performed in the $t$ integral. 
 $\Gamma(3) =2$ and $\Gamma(\frac{5}{2})= \frac{3}{4} \sqrt{\pi}$ are used in the last line. 
Integrating the above result, we can obtain the one-loop effective potential.\footnote{$\alpha$ indepedent constant 
which is an integration constant is irerevant in our analysis of the electroweak symmetry breaking.} 
\begin{equation}
 V_{\rm eff}^{g \ppp}
  = -\frac{1}{32 \pi^3} \sum_{n=1}^\infty 
  \left\{\sum_{m=1}^\infty 
  \frac{4}{\left((\pi R_5 n)^2 + (\pi R_6 m)^2\right)^3}
  + \frac{2}{(\pi R_5 n)^6}
  + \frac{3}{4(\pi R_5 n)^5 R_6}\right\}
  \cos(2 \pi n \alpha).
\end{equation}
The other potentials $V_{\rm eff}^{g\pmm}, V_{\rm eff}^{g\pmp}$ and $V_{\rm eff}^{g\mpm}$ can be calculated similarly. 
\begin{align}
  &V_{\rm eff}^{g \pmm}
  = -\frac{1}{16 \pi^3} \sum_{n=1}^\infty 
  \left\{\sum_{m=1}^\infty 
  \frac{4}{\left((\pi R_5 n)^2 + (\pi R_6 m)^2\right)^3}
  + \frac{2}{(\pi R_5 n)^6}
  + \frac{3}{4(\pi R_5 n)^5 R_6}\right\}
  \cos(\pi n \alpha),\\
  &V_{\rm eff}^{g \pmp}
  = -\frac{1}{16 \pi^3} \sum_{n=1}^\infty 
  \left\{\sum_{m=1}^\infty 
  \frac{4}{\left((\pi R_5 n)^2 + (\pi R_6 m)^2\right)^3}
  + \frac{2}{(\pi R_5 n)^6}
  + \frac{3}{4(\pi R_5 n)^5 R_6}\right\}
  (-1)^n \cos(\pi n \alpha),\\
  &V_{\rm eff}^{g \mpm}
  = -\frac{3}{16 \pi^3} \sum_{n=1}^\infty 
  \left\{\sum_{m=1}^\infty 
  \frac{4}{\left((\pi R_5 n)^2 + (\pi R_6 m)^2\right)^3}
  + \frac{2}{(\pi R_5 n)^6}
  - \frac{3}{4(\pi R_5 n)^5 R_6}\right\}
  (-1)^n \cos(\pi n \alpha)
\end{align}
Finally, the one-loop effective potential from the 4D gauge field contributions is obtained. 
\begin{align}
  V_{\rm eff}^g &= -\frac{1}{32 \pi^3} \sum_{n=1}^\infty
  \left\{\sum_{m=1}^\infty 
  \frac{4}{\left((\pi R_5 n)^2 + (\pi R_6 m)^2\right)^3}
  + \frac{2}{(\pi R_5 n)^6}
  + \frac{3}{4(\pi R_5 n)^5 R_6}\right\} \\
  & \quad \hspace{60mm} \times
  \left\{\cos(2 \pi n \alpha) + 2 \cos(\pi n \alpha)
  + 2(-1)^n \cos(\pi n \alpha)\right\} \notag \\
  & \quad 
  -\frac{1}{32 \pi^3} \sum_{n=1}^\infty 
  \left\{\sum_{m=1}^\infty 
  \frac{4}{\left((\pi R_5 n)^2 + (\pi R_6 m)^2\right)^3}
  + \frac{2}{(\pi R_5 n)^6}
  - \frac{3}{4(\pi R_5 n)^5 R_6}\right\}
  6(-1)^n \cos(\pi n \alpha).
\end{align}


\end{document}